\documentclass[11pt,a4paper]{article}
\usepackage{jheppub}
\usepackage{amsmath,amssymb}
\usepackage{booktabs}
\usepackage{placeins}
\usepackage{threeparttable}
\usepackage{graphicx}   % \resizebox in the below-target expression listing
\usepackage{tikz}
\usetikzlibrary{arrows.meta,positioning,shapes.geometric}
\usepackage{xcolor}
\usepackage{multirow}
\usepackage{orcidlink}

\usepackage[T1]{fontenc}
\hypersetup{colorlinks=true,linkcolor=blue,citecolor=blue,urlcolor=blue}
\usepackage[mathlines]{lineno}
\definecolor{ProofLine}{HTML}{595959}

\usepackage{xcolor}
\usepackage{bm}
\usepackage[arrows]{ezedits}
\definecolor{UMNRed}{RGB}{144,0,33}
\definecolor{StudentBlue}{RGB}{0,70,140}
\defineEditQuick{ZL}{UMNRed}
\defineEditQuick{YF}{StudentBlue}
\makeatletter
\let\ez@YFnote\YF
\DeclareRobustCommand{\YF}[1]{%
  \if@accept\ez@YFnote{#1}\else\if@reject\ez@YFnote{#1}\else\if@pdfcomm\ez@YFnote{#1}\else
    \YF@show{#1}\fi\fi\fi}
\newcommand{\YF@show}[1]{%
  \PackageWarning{ezedits}{Unresolved note by YF: "#1"}%
  \ifvmode\YF@block\fi
  \leavevmode
  \if@showarrows\marginnote{\textcolor{StudentBlue}{\Large$\bm{\Longleftarrow}$}}\fi
  \textcolor{StudentBlue}{\texttt{[YF: #1]}}%
  \YF@endblock}
\let\YF@endblock\relax
\def\YF@block{\begingroup\noindent\raggedright\def\YF@endblock{\par\endgroup}}
\makeatother

\newcommand{\ab}[2]{\langle #1\,#2\rangle}
\newcommand{\sq}[2]{[#1\,#2]}

\newcommand{\gbfs}{\textsc{gbfs}}
\newcommand{\gbfsraw}{\textsc{gbfs}\textsubscript{raw}}

\preprint{UMN-TH-4603/26}
\title{Fast, training-free simplification of spinor-helicity amplitudes}

\author[a]{Yifan Fei\,\orcidlink{0009-0009-8850-4767},}
\author[b]{Zhen Liu\,\orcidlink{0000-0002-3143-1976}}
\affiliation[a]{Department of Physics and Center for Field Theory and Particle Physics, Fudan University, Shanghai 200438, China}
\affiliation[b]{School of Physics and Astronomy, University of Minnesota, Minneapolis, MN 55455, USA}
\emailAdd{yffei24@m.fudan.edu.cn}
\emailAdd{zliuphys@umn.edu}

\abstract{We present a training-free method that simplifies massless spinor-helicity expressions by a greedy best-first search over on-shell identities, and test it at four, five, and six points. Applied to the color-ordered five-gluon MHV tree-level amplitude, the search derives the Parke--Taylor formula from all $1024$ Feynman-diagram expressions with different polarization reference spinors and up to $228$ numerator terms. On public test sets used by previous machine-learning methods, the search solves every case and generally outperforms these methods in solve rate. Its mean wall-clock time per case is $0.02$--$0.3\,\mathrm{s}$ on one CPU core. With longer searches and an alternative ranking of candidate expressions, it also solves all $21{,}600$ cases of a new suite of larger test expressions. Each result comes with an explicit derivation and is independently checked over finite fields at on-shell kinematics. For the classes studied here, the method effectively compresses long expressions produced by identity scrambling or Feynman-diagram calculations into compact forms. This is a first step toward extracting physics from amplitudes by relating distinct compact expressions, selecting representations with preferred factors and poles, and guiding the discovery of new amplitude structures.}

\begin{document}
\maketitle

\section{Introduction}
\label{sec:intro}
Scattering amplitudes in gauge and gravity theories admit remarkably compact representations in spinor-helicity variables~\cite{Parke:1986gb,Britto:2005fq,Dixon:1996wi,Dixon:2013uaa,Elvang:2015rqa,ArkaniHamed:2017jhn,Mangano:1990by,Cheung:2017pzi}. A Parke--Taylor $n$-gluon MHV amplitude, for example, is a single rational monomial in angle brackets. Diagrammatic and algebraic calculations rarely produce such forms directly, and they scale poorly with multiplicity unless color, helicity, and analytic structure are exploited to keep intermediate expressions under control~\cite{Dixon:1996wi,Dixon:2013uaa}. On-shell methods~\cite{Witten:2003nn,Cachazo:2004kj,Britto:2004ap,Britto:2005fq} reduce this growth but still leave the task of recognizing a compact form once a large equivalent expression has been produced, whether by Feynman diagrams, on-shell recursion, numerical reconstruction~\cite{Laurentis:2019bjh,DeLaurentis:2020qle,DeLaurentis:2022otd,Barrera:2025uin}, or generic computer algebra. Schouten identities, momentum conservation, and momentum-squared relations connect these equivalent forms, but each identity replaces a bracket with a sum of bracket monomials and can therefore enlarge an expression as well as shrink it.

Existing spinor-helicity packages provide tools for symbolic manipulation and simplification~\cite{Maitre:2007jq,AccettulliHuber:2023ldr}. Refs.~\cite{Cheung:2024CDS,Shih:2026SFT} approach expression simplification as a machine-learning problem. The public test sets of Ref.~\cite{Cheung:2024CDS}, also used in Ref.~\cite{Shih:2026SFT}, are generated by scrambling compact targets through random identity rewrites, and the task is to recover equivalent expressions no larger than those targets. Ref.~\cite{Shih:2026SFT} trains a supervised step-wise policy to imitate sequences of identity rewrites (hereafter SFT), while Ref.~\cite{Cheung:2024CDS} trains a one-shot sequence-to-sequence simplifier that maps a scrambled expression directly to a compact form (hereafter CDS). Both solve most cases of these sets. SFT returns an explicit derivation, whereas CDS returns only the final expression.

A fixed measure of expression size can also guide a search over identities without training. In dilogarithm simplification, Dersy, Schwartz, and Zhang reported solve rates of $91\%$ for a modified best-first search and $89\%$ for their reinforcement-learning agent with beam search~\cite{Dersy:2022bym}. A complementary line of research explores programmatic simplifications using coding agents driven by fixed numerical objectives~\cite{Gu:2026zbf}.

Denominators pose a further difficulty. Multi-particle invariants $s_{i_1\dots i_k}$ with $k>2$ and Dirac chains $\langle a|P|b]$ are sums of bracket monomials rather than single brackets. Such a factor can give rise to a physical pole or a spurious pole. Canceling a spurious denominator factor may require several identity rewrites.

In this paper we apply a training-free greedy best-first search (\gbfs) to spinor-helicity simplification, using the same multiplicity-dependent identity menu as SFT. The search ranks candidate expressions by a bracket-count objective. It keeps multi-particle invariants and Dirac chains in the denominator unexpanded and cancels each such factor once the numerator becomes divisible by it. Each output comes with an explicit derivation from the input.

We evaluate \gbfs\ on the public four-, five-, and six-point test sets of Ref.~\cite{Cheung:2024CDS} and on a size$\times$depth suite of larger inputs with more scrambling rewrites, generated by our reimplementation of its generator. \gbfs\ solves every case in the public sets, with a mean wall-clock time of $0.02$--$0.3\,\mathrm{s}$ per case on one CPU core. With longer searches and an alternative ranking of candidate expressions, it also solves the full size$\times$depth suite. On the five-point Yang--Mills set of Ref.~\cite{Shih:2026SFT}, whose $1024$ Feynman-diagram expressions differ only in the choice of polarization reference spinors and contain $8$ to $228$ numerator terms, \gbfs\ reduces every expression to the Parke--Taylor form. A synthetic example with a Dirac chain in the denominator shows how the search cancels a spurious pole by building the matching factor in the numerator.

The remainder of this paper is organized as follows. Section~\ref{sec:prelim} establishes the notation, defines the identity menu and bracket-count objective, and introduces the test sets and success criteria. Section~\ref{sec:method} describes the search algorithm, the alternative ranking, the treatment of sum-denominators, and the finite-field check. Section~\ref{sec:experiments} reports solve rates, wall-clock times, and search costs on the public sets, the size$\times$depth suite, and the five-point Yang--Mills set, and closes with the limits of the search. Section~\ref{sec:summary} summarizes the results and discusses extensions of the search.

\section{Basic setup}
\label{sec:prelim}
We first specify the spinor-helicity variables and identities used by the search, then define the measures of expression size and the benchmark success criteria.

We follow the conventions of Ref.~\cite{ArkaniHamed:2017jhn}. For each massless external momentum $p_i^\mu$, we introduce two-component spinors $\lambda_i$ and $\tilde\lambda_i$ satisfying $p_i^\mu\sigma_\mu=\lambda_i\tilde\lambda_i$. We define the Lorentz-invariant angle and square brackets,
\begin{equation}
  \ab{i}{j} \equiv \epsilon_{\alpha\beta}\lambda_i^\alpha\lambda_j^\beta, \qquad
  \sq{i}{j} \equiv \epsilon^{\dot\alpha\dot\beta}\tilde\lambda_{i\dot\alpha}\tilde\lambda_{j\dot\beta},
  \qquad s_{ij}=(p_i+p_j)^2=\ab{i}{j}\sq{j}{i}.
  \label{eq:brackets}
\end{equation}
Both brackets are antisymmetric. The amplitudes and intermediate expressions considered here are rational functions of these brackets. For symbolic manipulation, we write each bracket in canonical order $i<j$ and absorb the signs from antisymmetry into the coefficients.

The identity menu consists of three families of massless identities, each solved for a chosen bracket. The \emph{Schouten identity} expresses the linear dependence of any three two-component spinors. For four labels it reads
\begin{equation}
  \ab{i}{j}\ab{k}{l} = \ab{i}{k}\ab{j}{l} - \ab{i}{l}\ab{j}{k},
  \label{eq:schouten}
\end{equation}
and the same holds for square brackets. A Schouten move solves Eq.~\eqref{eq:schouten} for the chosen bracket $\ab{i}{j}$, with the spectator legs $k$ and $l$ drawn from the remaining $n-2$ legs.

The other two families are \emph{momentum conservation},
\begin{equation}
\sum_i\ab{a}{i}\sq{i}{b}=0,
\end{equation}
and the \emph{momentum-squared} relations $p_I^2=p_{\bar I}^2$, where $p_I$ is the sum of the momenta in a subset $I$ of the legs and $\bar I$ is its complement. For each canonical bracket, the menu contains $\binom{n-2}{2}$ Schouten, $2(n-1)$ momentum-conservation and $2^{n-2}$ momentum-squared variants. This gives $11$, $19$, and $32$ variants at $n=4$, $5$, and $6$, respectively. A move applies one variant either to every occurrence of the chosen bracket in the numerator or to one occurrence in a single numerator term.

An on-shell expression admits many algebraically equivalent forms. A physically simple form has a compact numerator and makes its poles and factorization channels manifest, as in the Parke--Taylor formula~\cite{Parke:1986gb}. As a computable measure of size, we use the factored bracket count, the number of brackets in the numerator counted with multiplicity, with the monomial factor common to all numerator terms counted only once. The raw numerator bracket count instead counts the brackets of each numerator term separately. For example, $\ab{1}{2}\ab{2}{3}\ab{3}{4}+\ab{1}{2}\ab{2}{3}\ab{4}{5}$ has a factored count of $4$ and a raw count of $6$. The objective of \gbfs\ is the ordered pair
\begin{equation}
  h=(\text{factored brackets},\,\text{terms}),
  \label{eq:search-objective}
\end{equation}
where the entries are the factored numerator bracket count and the number of additive numerator terms. We first compare the factored numerator bracket counts. If these are equal, we prefer the expression with fewer numerator terms. The raw count is used by the alternative frontier ordering in \S\ref{sec:gbfs} and by the success criteria below. Denominator factors that are sums of bracket monomials, such as multi-particle invariants and Dirac chains, are kept unexpanded, as described in \S\ref{sec:sumden}.

\noindent\textbf{Test sets and success criteria:} In the public sets and the size$\times$depth suite, each input is generated from a compact target by $K$ random scrambling moves, where $K$ is the generation depth. The task is to recover an expression representing the same amplitude as the input and no larger than the target under the criteria below. The public sets are the four-, five- and six-point test sets released by Ref.~\cite{Cheung:2024CDS} and also used for evaluation in Ref.~\cite{Shih:2026SFT}. The size$\times$depth suite of \S\ref{sec:suite} is generated by our reimplementation of the same generator. Besides the three identity families above, the generator can multiply a term by a unity factor such as $R/\ab{i}{j}$, where $R$ is a rewriting of $\ab{i}{j}$ by one identity of the menu, and add a multiple of the vanishing combination $R-\ab{i}{j}$. Such unity factors can introduce spurious poles. Starting from the target $0$, the generator must first add a vanishing combination. The public sets include cases generated in this way, which are labeled as vanishing.

On the public sets, a case is solved when both the raw numerator bracket count and the additive term count of the output are at most those of the target. On the size$\times$depth suite, a case is solved when the raw numerator bracket count of the output is at most that of the target.

\section{Method}
\label{sec:method}
We describe how the search explores identity rewrites, handles denominators containing sums, and checks that the resulting expressions preserve the input amplitude.

\subsection{Best-first identity search}\label{sec:gbfs}
\gbfs\ is a greedy best-first search~\cite{Doran:1966,Pearl:1984} over the identity menu of \S\ref{sec:prelim}, where each node represents an expression. After every move, the child is brought to a normal form by expanding the numerator into bracket monomials to collect like terms and canceling bracket powers common to the numerator and the monomial denominator, with sum-denominator factors handled as in \S\ref{sec:sumden}. All bracket counts refer to this normal form, allowing a single move to drastically shorten an expression when its terms collapse. At any given node, the algorithm enumerates all legal moves, scores each child by $h$ in Eq.~\eqref{eq:search-objective}, and retains new expressions on the frontier. The frontier contains all discovered expressions that have not yet been expanded. At each iteration, the node with the smallest $h$ is expanded and leaves the frontier, while the best expression found so far is retained separately. An expression whose normal form has already been reached is not added to the frontier again, which prevents cycles. Each node records the move that produced it, so the returned expression comes with an explicit derivation from the input. Every move preserves the expression's value.

A move can enlarge the expression by replacing a bracket with a sum of bracket monomials divided by a single bracket. Cancellations during normalization can reduce it, either in the same move or after further moves. Children with a higher $h$ than their parent remain on the frontier, allowing the search to follow derivations that first enlarge the expression and then reduce it.

\begin{figure}[t]
  \centering
  \includegraphics{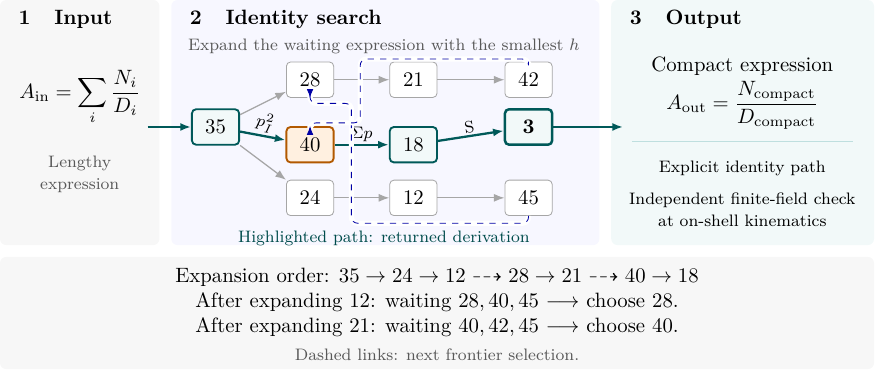}
  \caption{Best-first identity search, with schematic factored numerator bracket counts. Solid tree arrows are identity rewrites. Dashed links from $45$ to $28$ and from $42$ to $40$ indicate selection of the next waiting node. Neither $45$ nor $42$ is expanded. The separate sequence below gives the expansion order for the depicted candidates, assuming no earlier target stop. Expanding $12$ generates $45$, so the next waiting node is $28$; expanding $21$ generates $42$, so the next is $40$. The best expression found is retained even after its node has been expanded. The highlighted returned derivation $35\to40\to18\to3$ therefore differs from the expansion order and includes temporary growth. Edge labels denote Schouten ($\mathrm{S}$), momentum conservation ($\Sigma p$), and momentum-squared ($p_I^2$) identities. The output includes the derivation and an independent, probabilistic finite-field check performed after search. A finite budget need not reach a compact target or a globally minimal expression.}
  \label{fig:schematic}
\end{figure}

When the target is known, the search stops once a node satisfies the success criterion of \S\ref{sec:prelim}. Otherwise it returns the smallest node found when the move budget or the wall-clock time cap is exhausted (Fig.~\ref{fig:schematic}). The public sets and the size$\times$depth suite are evaluated in this target-stopped mode. In contrast, target-free runs are used for the Yang--Mills set of \S\ref{sec:ym5pt} and the sample of App.~\ref{app:below}, leaving the comparison against the target as a separate, post-run evaluation.

The search is bounded by a move budget, which counts every evaluated move, and a wall-clock time cap. On the public sets and the size$\times$depth suite we use $6\times10^{4}$ moves and a $300\,\mathrm{s}$ time cap per case. Because a target-stopped run ends as soon as the success criterion is met, it typically uses only a small fraction of this budget. Cases left unsolved at these limits are searched with a move budget of $2.4\times10^{5}$ and, if still unsolved, with $4\times10^{7}$ moves and a longer time cap. No run at these two move budgets reaches its time cap. Further extensions are described in \S\ref{sec:limits}. We refer to these searches as using increased budgets.

\gbfs\ ranks the frontier by $h$. We denote by \gbfsraw\ the variant that ranks it by the raw numerator bracket count alone. The identity menu and success criteria are unchanged. The two variants differ only in the order in which nodes are expanded, and \S\ref{sec:limits} shows that this order can substantially change the search cost of individual cases.

\FloatBarrier
\subsection{Sum-denominators}
\label{sec:sumden}
Physical amplitudes often contain multi-particle invariants $s_{i_1\dots i_k}$ with $k>2$ and Dirac-chain factors $\langle a|P|b]=\sum_{i\in P}\ab{a}{i}\sq{i}{b}$ in their denominators. Each such factor is a sum of bracket monomials, so the denominator is in general a bracket monomial multiplied by one or more such sums $S_i$. \gbfs\ expands Dirac chains and multi-particle invariants in the numerator into brackets but keeps each denominator sum $S_i$ unexpanded. A move rewrites only numerator brackets and leaves these denominator sums unchanged. During normalization after each move, any $S_i$ that divides the numerator as a polynomial in the brackets is canceled, and the quotient becomes the new numerator.

\begin{figure}[!htbp]
  \centering
  \includegraphics{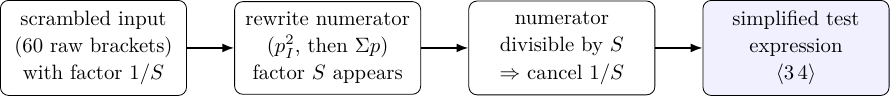}
  \caption{Cancellation of the Dirac-chain factor $S=\langle 2|(6{+}1)|5]$. \gbfs\ makes the numerator divisible by $S$ and cancels the common factors, reducing the synthetic input to the simplified test expression $\ab{3}{4}$. The full derivation is given in App.~\ref{app:sumden-example}.}
  \label{fig:worked}
\end{figure}

Removing a spurious factor $1/S_i$ therefore requires a derivation that makes the numerator divisible by $S_i$. Figure~\ref{fig:worked} illustrates this with a synthetic input containing a single sum-denominator factor, the six-point Dirac chain $S=\langle 2|(6{+}1)|5] =\ab{2}{6}\sq{6}{5}+\ab{2}{1}\sq{1}{5}$. Up to parity and an overall sign, this is the spurious denominator factor in the two-term BCFW representation of the split-helicity six-gluon NMHV amplitude~\cite{Britto:2004ap,Hodges:2009hk}.

Starting from an input with $60$ raw numerator brackets, \gbfs\ applies a momentum-squared move followed by a momentum-conservation move, after which the numerator is divisible by $S$. Canceling $S$ and the remaining common bracket factors reduces it to the simplified test expression $\ab{3}{4}$. This constructed example tests cancellation of the denominator factor. Keeping sum-denominators unexpanded allows \gbfs\ to simplify expressions with multi-particle invariants or Dirac chains in their denominators.

\FloatBarrier
\subsection{Randomized equality checks over finite fields}
\label{sec:modp}
Two expressions are equal if they coincide at every on-shell kinematic point where both are defined. We test this equality at random on-shell kinematic points over finite fields, a standard technique in amplitude computations~\cite{vonManteuffel:2014ixa,Peraro:2016wsq,Peraro:2019svx,Klappert:2019emp,DeLaurentis:2023qhd}. Following the polynomial view of Ref.~\cite{DeLaurentis:2022otd}, an expression is a rational function of the $4n$ spinor components $\lambda_i^{\alpha}$, $\tilde\lambda_i^{\dot\alpha}$, with each bracket of Eq.~\eqref{eq:brackets} a $2\times2$ minor. The on-shell kinematic points form the variety defined by momentum conservation,
\begin{equation}
  \sum_{i=1}^{n}\lambda_i^{\alpha}\,\tilde\lambda_i^{\dot\beta}=0 .
  \label{eq:momcons}
\end{equation}

We sample each kinematic point directly on this variety over $\mathbb{F}_p$. The components of $\lambda_1,\dots,\lambda_n$ and $\tilde\lambda_1,\dots,\tilde\lambda_{n-2}$ are drawn independently and uniformly from $\mathbb{F}_p$. For each value of $\dot\beta$, Eq.~\eqref{eq:momcons} gives a $2\times2$ linear system for $\tilde\lambda_{n-1}^{\dot\beta}$ and $\tilde\lambda_n^{\dot\beta}$, with determinant $\ab{n{-}1}{n}$. We discard points at which this determinant vanishes modulo $p$ and otherwise solve the two linear equations for the remaining spinor components. Since the brackets are constructed from explicit two-component spinors, they automatically satisfy the Schouten identity. Momentum conservation is imposed exactly, so the momentum-conservation and momentum-squared relations of \S\ref{sec:prelim} hold as well. Two equal expressions therefore take the same value at every sampled point where both are defined modulo $p$.

To compare $A$ and $B$, we write each expression as a single fraction, $A=N_A/D_A$ and $B=N_B/D_B$. Here $D_A$ and $D_B$ are the full denominators, including any sum-denominator factors $S_i$ of \S\ref{sec:sumden}. The cross-multiplied congruence
\begin{equation}
  N_A\,D_B \;\equiv\; N_B\,D_A \pmod{p}
  \label{eq:crossmult}
\end{equation}
is evaluated in $\mathbb{F}_p$, with each rational coefficient $a/b$ interpreted as $a\,b^{-1}$. We use $16$ independent kinematic points for each of the two primes $p_1=2^{31}-1$ and $p_2=2^{31}-19$. Points at which $D_A$ or $D_B$ vanishes modulo $p$ are discarded and resampled. The expressions $A$ and $B$ pass the finite-field check if Eq.~\eqref{eq:crossmult} holds at all $32$ valid points. If, for either prime, the resampling limit is reached before $16$ valid points are obtained, the comparison is inconclusive. Testing whether an expression vanishes identically amounts to setting $B=0$.

The probability of a false positive can be bounded using the Schwartz--Zippel lemma~\cite{Schwartz:1980,Zippel:1979}. After substituting the solutions of these linear systems and clearing denominators, the difference $N_AD_B-N_BD_A$ becomes a polynomial $Q_p$ over $\mathbb{F}_p$ in the $4n-4$ free spinor components. For an unequal pair, suppose that $Q_p$ is nonzero for both primes. Let $d_p$ be its total degree and $\alpha_p>0$ the probability that a sampled point is valid. Conditioned on validity, the probability that the point is a root of $Q_p$ is at most $\min(1,d_p/(\alpha_p p))$. Since the points are independent, the probability that the pair passes the finite-field check is bounded by
\begin{equation}
  \prod_{p\in\{p_1,p_2\}}
  \left[\min\!\left(1,\frac{d_p}{\alpha_p p}\right)\right]^{16}.
  \label{eq:conditional-modp-bound}
\end{equation}

We use the finite-field check to test whether the outputs of \gbfs, SFT, and CDS are equivalent to their respective inputs. We also checked every step of the \gbfs\ derivations behind Tables~\ref{tab:public}, \ref{tab:capdepth}, and~\ref{tab:capdepthext}, and all steps passed.

\section{Results}
\label{sec:experiments}
We evaluate \gbfs\ on the public sets, the size$\times$depth suite, and the five-point Yang--Mills set. On the public sets, we compare it with SFT, a supervised step-wise policy over the same identity menu~\cite{Shih:2026SFT}, and CDS, a sequence-to-sequence simplifier~\cite{Cheung:2024CDS}. On the size$\times$depth suite, we compare it with SFT. Each comparison uses identical inputs and the same success criterion.

\subsection{Public sets}
\label{sec:public}
Table~\ref{tab:public} reports the solve rate and the mean wall-clock time per case on the \emph{shallow} public test sets ($K\le3$; \S\ref{sec:prelim}). At a budget of $6 \times 10^{4}$ moves per case, \gbfs\ solves all four-, five-, and six-point expressions, with no case reaching the $300\,\mathrm{s}$ time cap. On a single CPU core, the mean wall-clock time per case is $0.021$, $0.088$, and $0.317\,\mathrm{s}$, respectively.

We use the inference settings of Ref.~\cite{Shih:2026SFT} for SFT and CDS, including their original stopping rules. SFT uses a $30\,\mathrm{s}$ wall-clock limit for each backtracking attempt, so its solve rate can vary with machine speed. In our runs, SFT leaves $10$, $38$, and $61$ cases unsolved at four, five, and six points, compared with $11$, $36$, and $63$ in Ref.~\cite{Shih:2026SFT}. After inference, we bring the saved SFT outputs and CDS candidates to the same normal form as the outputs of \gbfs\ and apply the common numerator-based success criterion of \S\ref{sec:prelim}. A CDS case is solved if the greedy output matches the target exactly or an equivalent candidate from beam search with $B=20$ meets this criterion.

The target is not a proven minimum, so an expression that meets the success criterion may still be simplified further. In a target-free run on $296$ cases labeled as non-vanishing in the five-point public set, \gbfs\ found expressions with fewer numerator brackets than their targets in $32$ cases. Appendix~\ref{app:below} lists these cases.

\begin{table}[t]
  \centering
  \normalsize
  \setlength{\tabcolsep}{2.5pt}
  \begin{tabular}{lccccccc}
    \toprule
    & & \multicolumn{2}{c}{\textbf{GBFS} ($6 \times 10^{4}$ moves)} & \multicolumn{2}{c}{SFT} & \multicolumn{2}{c}{CDS ($B{=}20$)} \\
    \cmidrule(lr){3-4}\cmidrule(lr){5-6}\cmidrule(lr){7-8}
    public set & $N$ & solve rate & mean s/case & solve rate & mean s/case & solve rate & mean s/case \\
    \midrule
    $4$-pt & $10020$ & $\mathbf{100.00}$ & {$0.021$} & {$99.90$} & {$1.56$} & {$98.24$} & {$0.167$} \\
    $5$-pt & $10270$ & $\mathbf{100.00}$ & {$0.088$} & {$99.63$} & {$3.77$} & {$95.99$} & {$0.164$} \\
    $6$-pt & $9934$  & $\mathbf{100.00}$ & {$0.317$} & {$99.39$} & {$6.56$} & {$96.94$} & {$0.165$} \\
    \bottomrule
  \end{tabular}
  \caption[Solve rate and wall-clock time on the public sets]{Solve rate (\%) and mean wall-clock time per case on the \emph{shallow} public test sets (\mbox{$K\le3$})~\cite{Shih:2026SFT,Cheung:2024CDS}. $N$ is the number of cases in each set. All methods use the target-relative success criterion of \S\ref{sec:prelim}, which requires the raw numerator bracket count and the additive term count of the output to be at most those of the target. We run \gbfs\ and SFT on a single CPU core and CDS on a single A800 GPU. Times are averaged over all cases. For SFT, the mean times over solved cases are $1.31$, $2.83$, and $5.14\,\mathrm{s}$ at four, five, and six points, respectively. \gbfs\ stops as soon as the success criterion is met. For CDS, timing covers the full call from the input expression to the returned result.}
  \label{tab:public}
\end{table}

\FloatBarrier
\subsection{The size$\times$depth suite}
\label{sec:suite}
The inputs in the public sets are generated with at most three scrambling moves, $K\le3$~\cite{Cheung:2024CDS}. To study larger and more deeply scrambled inputs, we evaluate \gbfs\ and SFT on the size$\times$depth suite, which varies the generation depth $K\in\{15,20,25,30\}$ and the bracket limit $L\in\{60,90,120,150,200,250\}$. Each input is obtained from its target by $K$ random scrambling moves. During generation, a move is rejected if the resulting expression has a raw numerator bracket count greater than $L$.

The suite contains $N=300$ cases in each $(n,K,L)$ cell. SFT uses its publicly released checkpoints, and the \gbfs\ move budget and time cap are those of \S\ref{sec:method}.

\begin{table}[t]
  \centering
  \normalsize
  \setlength{\tabcolsep}{4pt}
  \begin{tabular}{llcccccccc}
    \toprule
    & & \multicolumn{2}{c}{$L=60$} & \multicolumn{2}{c}{$L=90$} & \multicolumn{2}{c}{$L=120$} & \multicolumn{2}{c}{$L=150$} \\
    \cmidrule(lr){3-4}\cmidrule(lr){5-6}\cmidrule(lr){7-8}\cmidrule(lr){9-10}
    $n$ & $K$ & \textbf{GBFS}/$+$ & SFT & \textbf{GBFS}/$+$ & SFT & \textbf{GBFS}/$+$ & SFT & \textbf{GBFS}/$+$ & SFT \\
    \midrule
    \multirow{4}{*}{$4$} & $15$ & $\mathbf{100}$ & $98$ & $\mathbf{100}$ & $97$ & $\mathbf{100}$ & $97^{\dagger}$ & $99$/$\mathbf{100}$ & $90^{\dagger}$ \\
    & $20$ & $\mathbf{100}$ & $99$ & $\mathbf{100}$ & $99$ & $99$/$\mathbf{100}$ & $94^{\dagger}$ & $99$/$\mathbf{100}$ & $89^{\dagger}$ \\
    & $25$ & $\mathbf{100}$ & $99$ & $\mathbf{100}$ & $97$ & $99$/$\mathbf{100}$ & $96^{\dagger}$ & $99$/$99^{\ddagger}$ & $85^{\dagger}$ \\
    & $30$ & $\mathbf{100}$ & $98$ & $99$/$\mathbf{100}$ & $98$ & $\mathbf{100}$ & $93^{\dagger}$ & $\mathbf{100}$ & $90^{\dagger}$ \\
    \midrule
    \multirow{4}{*}{$5$} & $15$ & $\mathbf{100}$ & $\mathbf{100}$ & $\mathbf{100}$ & $99$ & $\mathbf{100}$ & $97$ & $\mathbf{100}$ & $89$ \\
    & $20$ & $\mathbf{100}$ & $99$ & $\mathbf{100}$ & $99$ & $\mathbf{100}$ & $96$ & $\mathbf{100}$ & $90$ \\
    & $25$ & $\mathbf{100}$ & $99$ & $\mathbf{100}$ & $98$ & $\mathbf{100}$ & $96$ & $99$/$\mathbf{100}$ & $92$ \\
    & $30$ & $\mathbf{100}$ & $\mathbf{100}$ & $\mathbf{100}$ & $99$ & $\mathbf{100}$ & $96$ & $\mathbf{100}$ & $90$ \\
    \midrule
    \multirow{4}{*}{$6$} & $15$ & $\mathbf{100}$ & $\mathbf{100}$ & $\mathbf{100}$ & $99$ & $\mathbf{100}$ & $97$ & $\mathbf{100}$ & $95$ \\
    & $20$ & $\mathbf{100}$ & $\mathbf{100}$ & $\mathbf{100}$ & $99$ & $\mathbf{100}$ & $97$ & $\mathbf{100}$ & $96$ \\
    & $25$ & $\mathbf{100}$ & $99$ & $\mathbf{100}$ & $99$ & $\mathbf{100}$ & $99$ & $\mathbf{100}$ & $96$ \\
    & $30$ & $\mathbf{100}$ & $99$ & $\mathbf{100}$ & $99$ & $\mathbf{100}$ & $97$ & $\mathbf{100}$ & $96$ \\
    \bottomrule
  \end{tabular}
  \caption{Solve rate (\%) on the size$\times$depth suite, $N=300$ per cell. Columns are the bracket limit $L$ and rows are the generation depth $K$. CDS was not evaluated on the full suite. The full suite, including $L=200$ and $250$, is given in Table~\ref{tab:capdepthext}. Rates are rounded to the nearest percent, except that a value below $100$ never rounds up to it. SFT uses the published checkpoints with its original inference configuration~\cite{Shih:2026SFT}. \gbfs\ entries $a$/$b$ report the solve rates at $6\times10^{4}$ moves and with increased budgets ($\gbfs^{+}$), respectively. A single value denotes identical rates. The $\ddagger$ marks the case unsolved by $\gbfs^{+}$ but solved by \gbfsraw\ (\S\ref{sec:limits}). Both methods are scored by the raw numerator bracket count relative to the target. $^{\dagger}$~The four-point SFT checkpoint encodes at most $8$ distinct brackets per term and can rewrite at most $10$ terms individually. The fractions of four-point inputs within both limits are $87\%$, $74\%$, $53\%$, and $33\%$ for $L=60$, $90$, $120$, and $150$, respectively. The marked entries therefore reflect both SFT's simplification procedure and the limitations of its published checkpoint. }
  \label{tab:capdepth}
\end{table}

\begin{figure}[tbp]
  \centering
  \includegraphics[width=\textwidth]{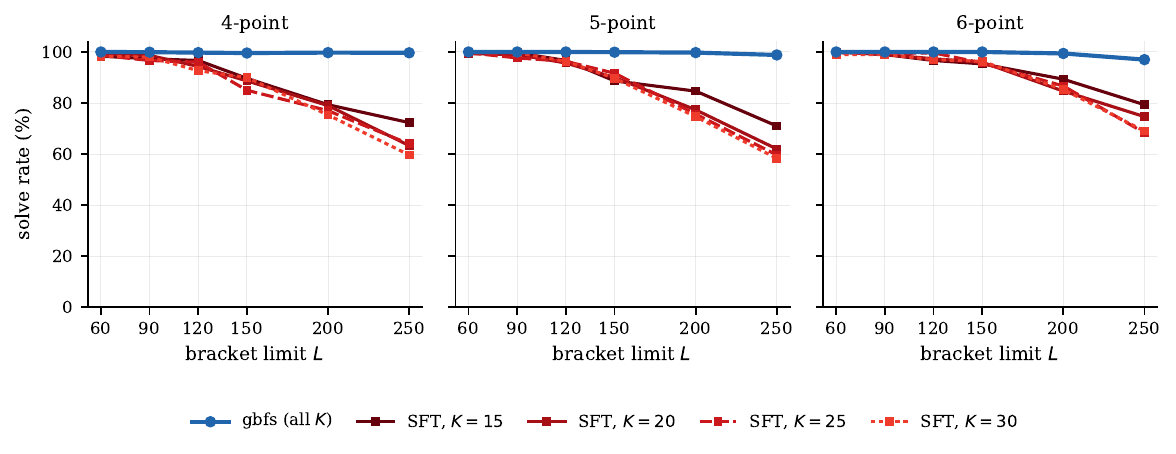}
  \caption{Solve rate versus bracket limit $L$. SFT is shown separately for each generation depth $K$ ($N=300$ cases per point), while \gbfs\ is averaged over the four values of $K$ ($N=1200$ cases per point) at a budget of $6\times10^{4}$ moves.}
  \label{fig:acccap}
\end{figure}

At the budget of $6\times10^{4}$ moves, \gbfs\ achieves a $100\%$ solve rate in $46$ out of $72$ cells and $93.7$--$99.7\%$ in the remaining $26$, with no case reaching the time cap. In contrast, at fixed $n$ and $K$, the absolute decrease in the SFT solve rate is $21\%$--$42\%$ as $L$ increases from $60$ to $250$. At fixed $n$ and $L$, the absolute variation across $K$ is at most $13\%$ (Fig.~\ref{fig:acccap}).

With increased budgets, \gbfs\ solves $21{,}599$ of the $21{,}600$ cases, and \gbfsraw\ solves the remaining one. We compare the two frontier orderings on this case in \S\ref{sec:limits}.

\begin{figure}[tbp]
  \centering
  \includegraphics[width=\textwidth]{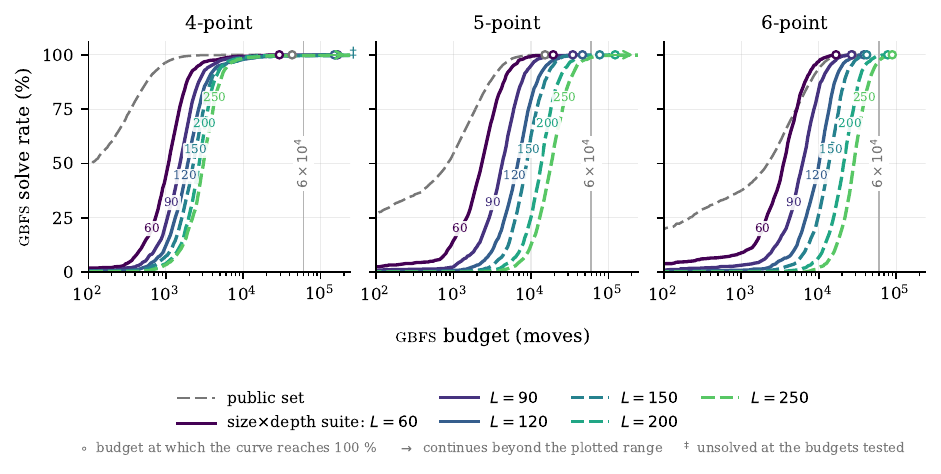}
  \caption{\gbfs\ solve rate versus the move budget. In each panel, colored curves show the size$\times$depth suite of Table~\ref{tab:budget}, one for each bracket limit $L$. The dashed gray curve shows the public set at the same $n$. The vertical line marks the budget of $6\times10^{4}$ moves used in Tables~\ref{tab:public} and~\ref{tab:capdepth}. Open circles mark the budget at which a curve reaches a $100\%$ solve rate, and arrows indicate that a curve continues beyond the plotted range. The four-point $L=150$ curve, marked $\ddagger$, does not reach $100\%$ within $4\times10^{9}$ moves.}
  \label{fig:budgetnpt}
\end{figure}

Table~\ref{tab:budget} and Fig.~\ref{fig:budgetnpt} show a steep increase in solve rate with the move budget. Table~\ref{tab:depthk} shows that, marginalized over $L$, the \gbfs\ solve rate is nearly flat in $K$, while the median number of evaluated moves varies by less than $20\%$ from $K=15$ to $K=30$.

\begin{table}[t]
  \centering
  \small
  \begin{tabular}{lccccrrrr}
    \toprule
    & \multicolumn{4}{c}{solve rate (\%)} & \multicolumn{4}{c}{median cost (moves)} \\
    \cmidrule(lr){2-5}\cmidrule(lr){6-9}
    $n$ & $K=15$ & $K=20$ & $K=25$ & $K=30$ & $K=15$ & $K=20$ & $K=25$ & $K=30$ \\
    \midrule
    $4$ & $99.83$ & $99.78$ & $99.56$ & $99.94$ & $1{,}997$ & $2{,}000$ & $1{,}996$ & $1{,}987$ \\
    $5$ & $99.89$ & $99.83$ & $99.61$ & $99.67$ & $7{,}052$ & $7{,}287$ & $7{,}987$ & $8{,}154$ \\
    $6$ & $99.89$ & $99.61$ & $99.33$ & $98.78$ & $10{,}437$ & $11{,}494$ & $11{,}874$ & $12{,}166$ \\
    \bottomrule
  \end{tabular}
  \caption{\gbfs\ solve rate and median search cost versus generation depth $K$, marginalized over the six bracket limits $L$, with $N=1{,}800$ cases per cell. The evaluation uses the budget of $6\times10^{4}$ moves and a $300\,\mathrm{s}$ time cap. The search cost is the number of evaluated moves, and its median is computed over cases solved within this budget.}
  
  \label{tab:depthk}
\end{table}

\begin{table}[t]
  \centering
  \small
  \begin{tabular}{llccccccc}
    \toprule
    & & \multicolumn{7}{c}{move budget} \\
    \cmidrule(lr){3-9}
    $n$ & $L$ & $10^{3}$ & $5\times10^{3}$ & $10^{4}$ & $3\times10^{4}$ & $6\times10^{4}$ & $10^{5}$ & $2\times10^{5}$ \\
    \midrule
    \multirow{7}{*}{$4$} & $60$ & $44.33$ & $98.08$ & $99.33$ & $\mathbf{100}$ & $\mathbf{100}$ & $\mathbf{100}$ & $\mathbf{100}$ \\
    & $90$ & $22.25$ & $97.33$ & $98.92$ & $99.83$ & $99.92$ & $99.92$ & $\mathbf{100}$ \\
    & $120$ & $12.00$ & $95.17$ & $97.92$ & $99.50$ & $99.75$ & $99.92$ & $\mathbf{100}$ \\
    & $150$ & $7.33$ & $93.67$ & $98.00$ & $99.42$ & $99.58$ & $99.67$ & $99.67$ \\
    & $200$ & $3.92$ & $93.75$ & $98.58$ & $99.67$ & $99.75$ & $99.75$ & $99.75$ \\
    & $250$ & $3.42$ & $87.67$ & $98.42$ & $99.50$ & $99.67$ & $99.67$ & $99.75$ \\
    & all & $15.54$ & $94.28$ & $98.53$ & $99.65$ & $99.78$ & $99.82$ & $99.86$ \\
    \midrule
    \multirow{7}{*}{$5$} & $60$ & $14.67$ & $91.83$ & $99.25$ & $\mathbf{100}$ & $\mathbf{100}$ & $\mathbf{100}$ & $\mathbf{100}$ \\
    & $90$ & $2.50$ & $62.25$ & $95.00$ & $99.75$ & $\mathbf{100}$ & $\mathbf{100}$ & $\mathbf{100}$ \\
    & $120$ & $0.83$ & $29.67$ & $79.58$ & $99.50$ & $\mathbf{100}$ & $\mathbf{100}$ & $\mathbf{100}$ \\
    & $150$ & $0.25$ & $11.58$ & $62.75$ & $99.50$ & $99.92$ & $\mathbf{100}$ & $\mathbf{100}$ \\
    & $200$ & $0.00$ & $4.08$ & $25.17$ & $95.42$ & $99.75$ & $99.92$ & $\mathbf{100}$ \\
    & $250$ & $0.00$ & $1.50$ & $12.50$ & $86.08$ & $98.83$ & $99.75$ & $99.83$ \\
    & all & $3.04$ & $33.49$ & $62.38$ & $96.71$ & $99.75$ & $99.94$ & $99.97$ \\
    \midrule
    \multirow{7}{*}{$6$} & $60$ & $8.92$ & $73.00$ & $96.92$ & $\mathbf{100}$ & $\mathbf{100}$ & $\mathbf{100}$ & $\mathbf{100}$ \\
    & $90$ & $3.08$ & $35.33$ & $84.50$ & $\mathbf{100}$ & $\mathbf{100}$ & $\mathbf{100}$ & $\mathbf{100}$ \\
    & $120$ & $0.92$ & $11.25$ & $49.25$ & $99.42$ & $\mathbf{100}$ & $\mathbf{100}$ & $\mathbf{100}$ \\
    & $150$ & $0.33$ & $4.17$ & $26.42$ & $97.67$ & $\mathbf{100}$ & $\mathbf{100}$ & $\mathbf{100}$ \\
    & $200$ & $0.08$ & $1.08$ & $6.75$ & $79.75$ & $99.42$ & $\mathbf{100}$ & $\mathbf{100}$ \\
    & $250$ & $0.00$ & $0.00$ & $2.08$ & $56.25$ & $97.00$ & $\mathbf{100}$ & $\mathbf{100}$ \\
    & all & $2.22$ & $20.81$ & $44.32$ & $88.85$ & $99.40$ & $\mathbf{100}$ & $\mathbf{100}$ \\
    \bottomrule
  \end{tabular}
  \caption{\gbfs\ solve rate versus the move budget on the size$\times$depth suite, marginalized over the four generation depths $K$, with $N=1{,}200$ cases in each $(n,L)$ row and $21{,}600$ cases in total. Each case is run with a budget of $2.4\times10^{5}$ moves, stopping once the success criterion is met. The recorded move counts determine the solve rates at the smaller budgets listed here. At $10^{4}$ moves, the six-point solve rate falls from $96.9\%$ at $L=60$ to $2.1\%$ at $L=250$, while the four-point rate falls only from $99.3\%$ to $98.4\%$.}
  \label{tab:budget}
\end{table}

\FloatBarrier
\subsection{Feynman-diagram gluon amplitudes at five points}
\label{sec:ym5pt}
The public sets and the size$\times$depth suite are constructed by scrambling compact target expressions. We next apply \gbfs\ to expressions obtained directly from Feynman diagrams.

Refs.~\cite{Cheung:2024CDS,Shih:2026SFT} introduced this five-gluon partial-amplitude set, which contains $1024$ expressions differing only in the assignment of polarization reference spinors to external legs. The expressions contain between $8$ and $228$ numerator terms, with a mean of about $90$. We use the common-sign release, in which $512$ of the original expressions have been multiplied by $-1$ to align their overall sign conventions. Every expression in this release is equal on shell to the same Parke--Taylor form~\cite{Parke:1986gb,Mangano:1990by},

\begin{equation}
  A_5^{\rm tree}(1^-2^-3^+4^+5^+) \;=\;
  \frac{\langle 12\rangle^{3}}{\langle 15\rangle\langle 23\rangle\langle 34\rangle\langle 45\rangle}\,,
  \label{eq:pt5}
\end{equation}
a single term of $3$ numerator brackets, or $7$ factors total including denominator multiplicity. Changing the reference spinors changes the form and size of the expression but not the amplitude, so the set varies the input size at a fixed target.

In target-free mode, \gbfs\ reduces all $1024$ cases to Eq.~\eqref{eq:pt5} with a common budget of $4\times10^{6}$ moves per case. Each run exhausts this budget without target-based stopping. The following search costs count moves up to the first discovery of the returned compact form, rather than the full run. The median search cost to reach this target is $9.64\times10^{4}$ moves, with a mean of $1.95\times10^{5}$.

The largest case, case~$1023$, has $228$ numerator terms and $2736$ brackets. \gbfs\ resolves this expression via an $18$-move derivation with $13$ momentum-conservation, $4$ Schouten, and $1$ momentum-squared moves, and the raw numerator bracket count decreases at every step ($2736 \to 1375 \to 1140 \to \cdots \to 8 \to 3$). Finding this derivation required evaluating $1.31\times10^{5}$ moves and took $49\,\mathrm{s}$ on a single CPU core or $7.6\,\mathrm{s}$ using $24$ threads.

Table~\ref{tab:ym5pt-cases} details this expression alongside two other representative cases. The timing columns come from separate runs that stop once the expression reaches the three numerator brackets of Eq.~\eqref{eq:pt5}. Stopping early does not change the frontier ordering, so these runs reach the target after the same number of moves as the target-free runs.

\begin{table}[t]
  \centering
  \begin{tabular}{lrrrrr}
    \toprule
    & & \multicolumn{2}{c}{moves} & \multicolumn{2}{c}{time to solution (s)} \\
    \cmidrule(lr){3-4}\cmidrule(lr){5-6}
    case & terms & derivation & search cost & $1$ core & $24$ threads \\
    \midrule
    $1023$ (largest)        & $228$ & $18$  & $1.31\times10^{5}$ & $49\,\mathrm{s}$  & $7.6\,\mathrm{s}$ \\
    $990$ (most demanding in \cite{Shih:2026SFT}) & $178$ & $34$  & $2.83\times10^{5}$ & $74\,\mathrm{s}$  & $12\,\mathrm{s}$ \\
    $935$ (most demanding here)  & $150$ & $108$ & $3.11\times10^{6}$ & $400\,\mathrm{s}$ & $73\,\mathrm{s}$ \\
    \bottomrule
  \end{tabular}
  \caption{Three representative cases from the five-point Yang--Mills set. Derivation length denotes the number of moves in the successful derivation, whereas the search cost reflects the total number of moves evaluated to find that derivation.}
  
  \label{tab:ym5pt-cases}
\end{table}

The median search cost increases with the initial numerator term count, from $8.0\times10^{3}$ moves in the $1$--$25$ term bin to $4.8\times10^{5}$ moves in the $201$--$228$ bin (Table~\ref{tab:ym5pt}). Within each bin, however, search costs vary by factors of approximately $7$ to $54$. Across bins, case~$935$ ($150$ terms) requires about $24$ times as many moves as the largest input, case~$1023$ ($228$ terms), and its derivation is $6$ times as long.

\newcommand{\ymFiveRows}{%
    $1$--$25$ & $47$ & $\mathbf{47/47}$ & $2.7$k & $8.0$k & $19.6$k & $7\times$ \\
    $26$--$50$ & $138$ & $\mathbf{138/138}$ & $7.4$k & $24.5$k & $337.6$k & $46\times$ \\
    $51$--$75$ & $206$ & $\mathbf{206/206}$ & $18.7$k & $56.2$k & $368.8$k & $20\times$ \\
    $76$--$100$ & $245$ & $\mathbf{245/245}$ & $29.6$k & $97.2$k & $977.2$k & $33\times$ \\
    $101$--$125$ & $184$ & $\mathbf{184/184}$ & $39.5$k & $198.8$k & $1,390.7$k & $35\times$ \\
    $126$--$150$ & $117$ & $\mathbf{117/117}$ & $57.9$k & $238.4$k & $3,112.2$k & $54\times$ \\
    $151$--$175$ & $50$ & $\mathbf{50/50}$ & $76.0$k & $351.8$k & $2,072.5$k & $27\times$ \\
    $176$--$200$ & $31$ & $\mathbf{31/31}$ & $96.1$k & $359.6$k & $1,455.9$k & $15\times$ \\
    $201$--$228$ & $6$ & $\mathbf{6/6}$ & $131.1$k & $477.6$k & $2,394.5$k & $18\times$ \\
    \midrule
    all & $1024$ & $\mathbf{1024/1024}$ & $2.7$k & $96.4$k & $3.11$M & $1166\times$ \\
}%
\begin{table}[t]
  \centering
  \begin{tabular}{lrrrrrr}
    \toprule
    & & & \multicolumn{3}{c}{search cost (moves)} & \\
    \cmidrule(lr){4-6}
    terms & $N$ & solved & min & median & max & spread \\
    \midrule
    \ymFiveRows
    \bottomrule
  \end{tabular}
  \caption{Solve rate and search cost for the $1024$ five-point Yang--Mills cases, binned by initial numerator term count. Spread is the ratio of the highest to the lowest search cost within each bin.}
  \label{tab:ym5pt}
\end{table}

\FloatBarrier
\subsection{Limits of the search}
\label{sec:limits}
With increased budgets, \gbfs\ leaves one four-point case ($K=25$, $L=150$) unsolved. Its target raw bracket count is $36$. \gbfs\ reaches a factored bracket count of $41$ after $14{,}494$ moves, but its minimum raw count remains $44$ after $4\times10^{9}$ moves. \gbfsraw\ solves the same case in $1.1\times10^{7}$ moves, returning an expression with $33$ raw brackets. For another four-point case ($K=15$, $L=250$), \gbfs\ requires $2.9\times10^{7}$ moves, compared with $4{,}738$ for \gbfsraw.

At a fixed budget of $2.4\times10^{5}$ moves, the solve rate of \gbfs\ exceeds that of \gbfsraw\ by approximately $0.06\%$ in absolute terms. With increased budgets, \gbfsraw\ solves the entire suite. The two orderings trade performance across cases. We retain the factored count as the default because it matches the objective in Eq.~\eqref{eq:search-objective} and solves slightly more cases at the fixed budget.  The raw ordering provides an alternative for the hard tail. These results come from separate runs of the two orderings and do not test a rule that switches between them automatically. More generally, different counting rules favor different expression forms. Designing objectives that favor specified factors or pole structures is a direction for future components beyond the compression algorithm studied here.

In the exceptional $4\times10^{9}$-move search above, many generated candidates remain waiting for expansion, making the frontier costly to hold in memory. A candidate is generated when an identity move produces it and expanded when it is selected to generate further candidates. The frontier could be stored on disk to reduce RAM use at the cost of additional input/output operations. It could also be bounded in size, which may change which solutions are found within a fixed budget.

\FloatBarrier
\section{Conclusion and outlook}
\label{sec:summary}
Recovering a compact form from a scrambled spinor-helicity expression can be posed as a search over sequences of identity moves. \gbfs\ solves every case in the public sets at a budget of $6\times10^{4}$ moves, exceeding the solve rates of SFT and CDS. With increased budgets, it solves $21{,}599$ of the $21{,}600$ cases in the size$\times$depth suite, and \gbfsraw\ solves the remaining one. In target-free mode, \gbfs\ reduces all $1024$ Feynman-diagram expressions of the five-point Yang--Mills set to the Parke--Taylor form. Keeping sum-denominators unexpanded allows a spurious denominator factor to cancel once the numerator becomes divisible by it. Each output comes with an explicit derivation and is compared with its input using the finite-field check. The six-point example in App.~\ref{app:case144} gives a $13$-move derivation reducing $115$ raw numerator brackets to $1$, illustrating how the returned trace makes a multi-step simplification explicit.

Target-free search can also improve upon the supplied reference forms. Among $296$ sampled five-point cases, it finds $32$ outputs with fewer raw numerator brackets than their targets (App.~\ref{app:below}), showing that these targets need not be the simplest representations.

The cost of expanding a node depends on the number of legal moves, which is determined by the identity menu and the brackets and terms in the numerator. The menu is generated for arbitrary $n$ and contains $11$ identity variants per canonical bracket at four points and $32$ at six. Evaluating the search at seven and eight points is a natural next step. Exceptionally long searches can also accumulate a large frontier, as illustrated by the four-point case in \S\ref{sec:limits}. Hybrid guidance~\cite{vonHippel:2025okr,Song:2025pwy,Zeng:2025xbh,Shih:2026jfe,Berman:2026jgq} offers one way to reduce the total search time and memory usage. A learned model could rank frontier nodes to reduce the number of expansions needed to reach a compact expression.

A further extension concerns Dirac chains with momentum sums, such as $\langle a|(p_1+p_3)|b]$, and multi-particle invariants, which are currently expanded into brackets when they appear in a numerator. Identities that act on these objects directly would let the search keep their compact form.

In a preliminary six-point NMHV test, target-free search reduces the expanded numerator from 126 to 32 terms while retaining the input denominator. A systematic study of six- to eight-point NMHV tree amplitudes and, subsequently, rational parts of one-loop amplitudes would test the method on more complex compact targets.

A longer-term goal is a single simplifier for massless and massive spinor-helicity amplitudes. Massive spinors carry little-group indices and satisfy mass-dependent identities~\cite{ArkaniHamed:2017jhn}, and on-shell recursion provides compact massive amplitudes against which such a simplifier can be tested~\cite{Ema:2024ALT,Ema:2024EW,Ema:2025Compton}. Extending the symbolic representation to massive spinors and adding the corresponding identities to the menu would allow the same search algorithm to be applied to massive amplitudes.

We call the compressor presented here \texttt{AmpZip}, the first component of \texttt{AmpLens}, a developing toolkit for exposing the analytic structure of scattering amplitudes. Planned components will address extraction of prescribed prefactors, residue extraction at specified poles, and machine-learning-informed search between compact expressions for selected high-value targets. These capabilities remain future work beyond the compression results reported here. The code, input data and evaluation commands used in this paper will be released through \href{https://github.com/AmpLens/AmpLens}{GitHub} with the arXiv submission.

\acknowledgments
We thank David Shih, Lance Dixon, and Jiayin Gu for helpful conversations, and David Shih and Aur\'elien Dersy for providing the five-point Yang--Mills expressions used in this work. Y.F. is supported by the National Natural Science Foundation of China under Grant No. 12375091 at Fudan University. Z.L. is supported by the Department of Energy under Grant No. DE-SC0011842 and in part by a Sloan Research Fellowship from the Alfred P. Sloan Foundation at the University of Minnesota. This work was assisted, in particular in the programming and in running the computations, by Anthropic's Claude and OpenAI's GPT models.

\appendix
\newcommand{\exSumdenInput}{%
\begin{equation*}\mathcal{E}_{0}=\frac{N_{0}}{\ab{3}{5}\sq{2}{6}\,S},\qquad S=\langle 2|(6{+}1)|5]=\ab{2}{6}\sq{6}{5} + \ab{2}{1}\sq{1}{5},\end{equation*}
}%
\newcommand{\exSumdenNumerator}{%
{\small
\begin{align*}
  N_{0} ={}& \underline{\ab{1}{2}}\ab{3}{4}\ab{3}{5}\sq{1}{2}\sq{5}{6} - \ab{1}{2}\ab{3}{4}\ab{3}{5}\sq{1}{5}\sq{2}{6} \\
  & + \ab{1}{3}\ab{3}{4}\ab{4}{5}\sq{1}{6}\sq{4}{5} + \ab{1}{4}\ab{3}{4}\ab{3}{5}\sq{1}{4}\sq{5}{6} \\
  & + \ab{1}{5}\ab{3}{4}\ab{3}{5}\sq{1}{5}\sq{5}{6} + \ab{1}{6}\ab{3}{4}\ab{3}{5}\sq{1}{6}\sq{5}{6} \\
  & + \ab{2}{3}\ab{3}{4}\ab{4}{5}\sq{2}{6}\sq{4}{5} + \ab{2}{4}\ab{3}{4}\ab{3}{5}\sq{2}{4}\sq{5}{6} \\
  & + \ab{2}{5}\ab{3}{4}\ab{3}{5}\sq{2}{5}\sq{5}{6} + \ab{3}{4}\ab{3}{5}\ab{4}{6}\sq{4}{6}\sq{5}{6} \\
  & + \ab{3}{4}\ab{3}{5}\ab{5}{6}\sq{5}{6}^{2} - \ab{3}{4}^{2}\ab{4}{5}\sq{4}{5}\sq{4}{6}.
\end{align*}}
}%
\newcommand{\exSumdenStepRows}{%
  $0$ & --- & input & $60$ & $49$ & $12$ \\
  $1^{\hyperlink{app:sum-denominator-star}{*}}$ & momentum-squared & $\ab{1}{2} \to \text{(R1)}$ & $30$ & $25$ & $6$ \\
  $2$ & momentum conservation & $\ab{1}{3} \to \bigl(- \ab{2}{3}\sq{2}{6} + \ab{3}{4}\sq{4}{6} + \ab{3}{5}\sq{5}{6}\bigr)/\sq{1}{6}$ & $1$ & $1$ & $1$ \\
}%
\newcommand{\exSumdenWideRules}{%
\noindent(R1)\quad $\ab{1}{2}\to N/\sq{1}{2}$ with
{\small
\begin{align*}
  N ={}& -\ab{1}{4}\sq{1}{4} - \ab{1}{5}\sq{1}{5} - \ab{1}{6}\sq{1}{6} - \ab{2}{4}\sq{2}{4} - \ab{2}{5}\sq{2}{5} - \ab{2}{6}\sq{2}{6} \\
  & - \ab{4}{5}\sq{4}{5} - \ab{4}{6}\sq{4}{6} - \ab{5}{6}\sq{5}{6}.
\end{align*}}
}%
\newcommand{\exSumdenSteps}{%
\par\smallskip\noindent\textbf{Step 1}\nobreak
{\small
\begin{align*}
  N_{1} ={}& -\ab{1}{2}\ab{3}{4}\ab{3}{5}\sq{1}{5}\sq{2}{6} + \underline{\ab{1}{3}}\ab{3}{4}\ab{4}{5}\sq{1}{6}\sq{4}{5} \\
  & + \ab{2}{3}\ab{3}{4}\ab{4}{5}\sq{2}{6}\sq{4}{5} - \ab{2}{6}\ab{3}{4}\ab{3}{5}\sq{2}{6}\sq{5}{6} \\
  & - \ab{3}{4}\ab{3}{5}\ab{4}{5}\sq{4}{5}\sq{5}{6} - \ab{3}{4}^{2}\ab{4}{5}\sq{4}{5}\sq{4}{6}
\end{align*}}
\noindent with $\mathcal{E}_{1}=N_{1}/(\ab{3}{5}\sq{2}{6}\,S)$.\par
\par\smallskip\noindent\textbf{Step 2}\nobreak
\begin{equation*}\mathcal{E}_{2} = \frac{\ab{3}{4}\ab{3}{5}\sq{2}{6}\,S}{\ab{3}{5}\sq{2}{6}\,S} = \ab{3}{4}.\end{equation*}
}%
\newcommand{\exDeepStepRows}{%
  $0$ & --- & input & $115$ & $115$ & $23$ \\
  $1^{\hyperlink{app:case144-star}{*}}$ & Schouten & $\sq{2}{3} \to \bigl(\sq{2}{4}\sq{3}{5} - \sq{2}{5}\sq{3}{4}\bigr)/\sq{4}{5}$ & $100$ & $100$ & $20$ \\
  $2^{\hyperlink{app:case144-star}{*}}$ & Schouten & $\ab{1}{5} \to \bigl(\ab{1}{4}\ab{5}{6} + \ab{1}{6}\ab{4}{5}\bigr)/\ab{4}{6}$ & $90$ & $90$ & $18$ \\
  $3$ & Schouten & $\ab{1}{3} \to \bigl(\ab{1}{4}\ab{3}{5} - \ab{1}{5}\ab{3}{4}\bigr)/\ab{4}{5}$ & $68$ & $68$ & $17$ \\
  $4$ & Schouten & $\sq{1}{2} \to \bigl(\sq{1}{3}\sq{2}{4} - \sq{1}{4}\sq{2}{3}\bigr)/\sq{3}{4}$ & $64$ & $64$ & $16$ \\
  $5^{\hyperlink{app:case144-star}{*}}$ & Schouten & $\sq{1}{3} \to \bigl(\sq{1}{5}\sq{3}{6} - \sq{1}{6}\sq{3}{5}\bigr)/\sq{5}{6}$ & $60$ & $60$ & $15$ \\
  $6^{\hyperlink{app:case144-star}{*}}$ & Schouten & $\sq{1}{3} \to \bigl(\sq{1}{4}\sq{3}{5} - \sq{1}{5}\sq{3}{4}\bigr)/\sq{4}{5}$ & $56$ & $56$ & $14$ \\
  $7^{\hyperlink{app:case144-star}{*}}$ & Schouten & $\sq{1}{4} \to \bigl(\sq{1}{5}\sq{4}{6} - \sq{1}{6}\sq{4}{5}\bigr)/\sq{5}{6}$ & $52$ & $52$ & $13$ \\
  $8$ & momentum conservation & $\ab{3}{5} \to \bigl(- \ab{1}{5}\sq{1}{4} - \ab{2}{5}\sq{2}{4} - \ab{5}{6}\sq{4}{6}\bigr)/\sq{3}{4}$ & $48$ & $48$ & $12$ \\
  $9^{\hyperlink{app:case144-star}{*}}$ & momentum conservation & $\ab{2}{3} \to \bigl(- \ab{2}{4}\sq{1}{4} - \ab{2}{5}\sq{1}{5} - \ab{2}{6}\sq{1}{6}\bigr)/\sq{1}{3}$ & $44$ & $44$ & $11$ \\
  $10$ & Schouten & $\sq{2}{4} \to \bigl(\sq{2}{3}\sq{4}{5} + \sq{2}{5}\sq{3}{4}\bigr)/\sq{3}{5}$ & $28$ & $28$ & $7$ \\
  $11$ & momentum conservation & $\ab{2}{3} \to \bigl(\ab{1}{2}\sq{1}{4} + \ab{2}{5}\sq{4}{5} + \ab{2}{6}\sq{4}{6}\bigr)/\sq{3}{4}$ & $15$ & $15$ & $5$ \\
  $12$ & momentum conservation & $\ab{1}{2} \to \bigl(\ab{1}{4}\sq{3}{4} + \ab{1}{5}\sq{3}{5} + \ab{1}{6}\sq{3}{6}\bigr)/\sq{2}{3}$ & $9$ & $7$ & $3$ \\
  $13$ & Schouten & $\sq{1}{3} \to \bigl(\sq{1}{4}\sq{3}{5} - \sq{1}{5}\sq{3}{4}\bigr)/\sq{4}{5}$ & $1$ & $1$ & $1$ \\
}%
\newcommand{\exDeepWideRules}{%
}%
\newcommand{\exDeepTarget}{%
\begin{equation*}\frac{- 3\ab{1}{4}}{\ab{2}{4}^{2}\ab{3}{6}^{2}\sq{2}{4}\sq{2}{5}\sq{3}{5}}.\end{equation*}
}%
\newcommand{\exDeepSteps}{%
\par\smallskip\noindent\textbf{Step 0}\nobreak
{\small
\begin{align*}
  N_{0} ={}& 15\ab{1}{3}\ab{4}{5}\sq{1}{5}\sq{3}{4}\sq{3}{5} + 12\ab{1}{4}\ab{1}{5}\sq{1}{4}\sq{1}{5}\sq{3}{5} \\
  & - 3\ab{1}{4}\ab{1}{6}\sq{1}{3}\sq{1}{4}\sq{5}{6} - 3\ab{1}{4}\ab{1}{6}\sq{1}{4}\sq{1}{6}\sq{3}{5} \\
  & - 12\ab{1}{4}\ab{2}{3}\sq{1}{2}\sq{3}{4}\sq{3}{5} - 12\ab{1}{4}\ab{2}{3}\sq{1}{4}\sq{2}{3}\sq{3}{5} \\
  & + 3\ab{1}{4}\ab{2}{3}\sq{1}{5}\sq{2}{3}\sq{3}{4} - 12\ab{1}{4}\ab{2}{4}\sq{1}{4}\sq{2}{3}\sq{4}{5} \\
  & - 12\ab{1}{4}\ab{2}{4}\sq{1}{4}\sq{2}{5}\sq{3}{4} - 3\ab{1}{4}\ab{2}{5}\sq{1}{3}\sq{2}{5}\sq{4}{5} \\
  & + 3\ab{1}{4}\ab{2}{5}\sq{1}{4}\sq{2}{5}\sq{3}{5} - 3\ab{1}{4}\ab{2}{6}\sq{1}{5}\sq{2}{3}\sq{4}{6} \\
  & - 12\ab{1}{4}\ab{2}{6}\sq{1}{6}\sq{2}{4}\sq{3}{5} + 12\ab{1}{4}\ab{5}{6}\sq{1}{4}\sq{3}{5}\sq{5}{6} \\
  & + 15\ab{1}{4}\ab{5}{6}\sq{1}{6}\sq{3}{5}\sq{4}{5} + 3\ab{1}{4}^{2}\sq{1}{3}\sq{1}{4}\sq{4}{5} \\
  & - 6\ab{1}{4}^{2}\sq{1}{4}^{2}\sq{3}{5} + 15\ab{1}{5}\ab{3}{4}\sq{1}{5}\sq{3}{4}\sq{3}{5} \\
  & + 3\ab{1}{5}\ab{4}{6}\sq{1}{4}\sq{3}{5}\sq{5}{6} - 3\ab{1}{6}\ab{4}{5}\sq{1}{4}\sq{3}{5}\sq{5}{6} \\
  & + 3\ab{2}{6}\ab{4}{5}\underline{\sq{2}{3}}\sq{4}{5}\sq{5}{6} - 3\ab{2}{6}\ab{4}{5}\sq{2}{4}\sq{3}{5}\sq{5}{6} \\
  & + 3\ab{2}{6}\ab{4}{5}\sq{2}{5}\sq{3}{4}\sq{5}{6}
\end{align*}}
\noindent with $\mathcal{E}_{0}=N_{0}/(\ab{1}{4}\ab{2}{4}^{2}\ab{3}{6}^{2}\sq{1}{4}^{2}\sq{2}{4}\sq{2}{5}\sq{3}{5}^{2})$.\par
\par\smallskip\noindent\textbf{Step 1}\nobreak
{\small
\begin{align*}
  N_{1} ={}& 15\ab{1}{3}\ab{4}{5}\sq{1}{5}\sq{3}{4}\sq{3}{5} + 12\ab{1}{4}\ab{1}{5}\sq{1}{4}\sq{1}{5}\sq{3}{5} \\
  & - 3\ab{1}{4}\ab{1}{6}\sq{1}{3}\sq{1}{4}\sq{5}{6} - 3\ab{1}{4}\ab{1}{6}\sq{1}{4}\sq{1}{6}\sq{3}{5} \\
  & - 12\ab{1}{4}\ab{2}{3}\sq{1}{2}\sq{3}{4}\sq{3}{5} - 12\ab{1}{4}\ab{2}{3}\sq{1}{4}\sq{2}{3}\sq{3}{5} \\
  & + 3\ab{1}{4}\ab{2}{3}\sq{1}{5}\sq{2}{3}\sq{3}{4} - 12\ab{1}{4}\ab{2}{4}\sq{1}{4}\sq{2}{3}\sq{4}{5} \\
  & - 12\ab{1}{4}\ab{2}{4}\sq{1}{4}\sq{2}{5}\sq{3}{4} - 3\ab{1}{4}\ab{2}{5}\sq{1}{3}\sq{2}{5}\sq{4}{5} \\
  & + 3\ab{1}{4}\ab{2}{5}\sq{1}{4}\sq{2}{5}\sq{3}{5} - 3\ab{1}{4}\ab{2}{6}\sq{1}{5}\sq{2}{3}\sq{4}{6} \\
  & - 12\ab{1}{4}\ab{2}{6}\sq{1}{6}\sq{2}{4}\sq{3}{5} + 12\ab{1}{4}\ab{5}{6}\sq{1}{4}\sq{3}{5}\sq{5}{6} \\
  & + 15\ab{1}{4}\ab{5}{6}\sq{1}{6}\sq{3}{5}\sq{4}{5} + 3\ab{1}{4}^{2}\sq{1}{3}\sq{1}{4}\sq{4}{5} \\
  & - 6\ab{1}{4}^{2}\sq{1}{4}^{2}\sq{3}{5} + 15\ab{1}{5}\ab{3}{4}\sq{1}{5}\sq{3}{4}\sq{3}{5} \\
  & + 3\underline{\ab{1}{5}}\ab{4}{6}\sq{1}{4}\sq{3}{5}\sq{5}{6} - 3\ab{1}{6}\ab{4}{5}\sq{1}{4}\sq{3}{5}\sq{5}{6}
\end{align*}}
\noindent with $\mathcal{E}_{1}=N_{1}/(\ab{1}{4}\ab{2}{4}^{2}\ab{3}{6}^{2}\sq{1}{4}^{2}\sq{2}{4}\sq{2}{5}\sq{3}{5}^{2})$.\par
\par\smallskip\noindent\textbf{Step 2}\nobreak
{\small
\begin{align*}
  N_{2} ={}& 15\underline{\ab{1}{3}}\ab{4}{5}\sq{1}{5}\sq{3}{4}\sq{3}{5} + 12\ab{1}{4}\ab{1}{5}\sq{1}{4}\sq{1}{5}\sq{3}{5} \\
  & - 3\ab{1}{4}\ab{1}{6}\sq{1}{3}\sq{1}{4}\sq{5}{6} - 3\ab{1}{4}\ab{1}{6}\sq{1}{4}\sq{1}{6}\sq{3}{5} \\
  & - 12\ab{1}{4}\ab{2}{3}\sq{1}{2}\sq{3}{4}\sq{3}{5} - 12\ab{1}{4}\ab{2}{3}\sq{1}{4}\sq{2}{3}\sq{3}{5} \\
  & + 3\ab{1}{4}\ab{2}{3}\sq{1}{5}\sq{2}{3}\sq{3}{4} - 12\ab{1}{4}\ab{2}{4}\sq{1}{4}\sq{2}{3}\sq{4}{5} \\
  & - 12\ab{1}{4}\ab{2}{4}\sq{1}{4}\sq{2}{5}\sq{3}{4} - 3\ab{1}{4}\ab{2}{5}\sq{1}{3}\sq{2}{5}\sq{4}{5} \\
  & + 3\ab{1}{4}\ab{2}{5}\sq{1}{4}\sq{2}{5}\sq{3}{5} - 3\ab{1}{4}\ab{2}{6}\sq{1}{5}\sq{2}{3}\sq{4}{6} \\
  & - 12\ab{1}{4}\ab{2}{6}\sq{1}{6}\sq{2}{4}\sq{3}{5} + 15\ab{1}{4}\ab{5}{6}\sq{1}{4}\sq{3}{5}\sq{5}{6} \\
  & + 15\ab{1}{4}\ab{5}{6}\sq{1}{6}\sq{3}{5}\sq{4}{5} + 3\ab{1}{4}^{2}\sq{1}{3}\sq{1}{4}\sq{4}{5} \\
  & - 6\ab{1}{4}^{2}\sq{1}{4}^{2}\sq{3}{5} + 15\ab{1}{5}\ab{3}{4}\sq{1}{5}\sq{3}{4}\sq{3}{5}
\end{align*}}
\noindent with $\mathcal{E}_{2}=N_{2}/(\ab{1}{4}\ab{2}{4}^{2}\ab{3}{6}^{2}\sq{1}{4}^{2}\sq{2}{4}\sq{2}{5}\sq{3}{5}^{2})$.\par
\par\smallskip\noindent\textbf{Step 3}\nobreak
{\small
\begin{align*}
  N_{3} ={}& 3\ab{1}{4}\sq{1}{3}\sq{1}{4}\sq{4}{5} - 6\ab{1}{4}\sq{1}{4}^{2}\sq{3}{5} + 12\ab{1}{5}\sq{1}{4}\sq{1}{5}\sq{3}{5} \\
  & - 3\ab{1}{6}\sq{1}{3}\sq{1}{4}\sq{5}{6} - 3\ab{1}{6}\sq{1}{4}\sq{1}{6}\sq{3}{5} - 12\ab{2}{3}\underline{\sq{1}{2}}\sq{3}{4}\sq{3}{5} \\
  & - 12\ab{2}{3}\sq{1}{4}\sq{2}{3}\sq{3}{5} + 3\ab{2}{3}\sq{1}{5}\sq{2}{3}\sq{3}{4} - 12\ab{2}{4}\sq{1}{4}\sq{2}{3}\sq{4}{5} \\
  & - 12\ab{2}{4}\sq{1}{4}\sq{2}{5}\sq{3}{4} - 3\ab{2}{5}\sq{1}{3}\sq{2}{5}\sq{4}{5} + 3\ab{2}{5}\sq{1}{4}\sq{2}{5}\sq{3}{5} \\
  & - 3\ab{2}{6}\sq{1}{5}\sq{2}{3}\sq{4}{6} - 12\ab{2}{6}\sq{1}{6}\sq{2}{4}\sq{3}{5} + 15\ab{3}{5}\sq{1}{5}\sq{3}{4}\sq{3}{5} \\
  & + 15\ab{5}{6}\sq{1}{4}\sq{3}{5}\sq{5}{6} + 15\ab{5}{6}\sq{1}{6}\sq{3}{5}\sq{4}{5}
\end{align*}}
\noindent with $\mathcal{E}_{3}=N_{3}/(\ab{2}{4}^{2}\ab{3}{6}^{2}\sq{1}{4}^{2}\sq{2}{4}\sq{2}{5}\sq{3}{5}^{2})$.\par
\par\smallskip\noindent\textbf{Step 4}\nobreak
{\small
\begin{align*}
  N_{4} ={}& 3\ab{1}{4}\sq{1}{3}\sq{1}{4}\sq{4}{5} - 6\ab{1}{4}\sq{1}{4}^{2}\sq{3}{5} + 12\ab{1}{5}\sq{1}{4}\sq{1}{5}\sq{3}{5} \\
  & - 3\ab{1}{6}\underline{\sq{1}{3}}\sq{1}{4}\sq{5}{6} - 3\ab{1}{6}\sq{1}{4}\sq{1}{6}\sq{3}{5} - 12\ab{2}{3}\sq{1}{3}\sq{2}{4}\sq{3}{5} \\
  & + 3\ab{2}{3}\sq{1}{5}\sq{2}{3}\sq{3}{4} - 12\ab{2}{4}\sq{1}{4}\sq{2}{3}\sq{4}{5} - 12\ab{2}{4}\sq{1}{4}\sq{2}{5}\sq{3}{4} \\
  & - 3\ab{2}{5}\sq{1}{3}\sq{2}{5}\sq{4}{5} + 3\ab{2}{5}\sq{1}{4}\sq{2}{5}\sq{3}{5} - 3\ab{2}{6}\sq{1}{5}\sq{2}{3}\sq{4}{6} \\
  & - 12\ab{2}{6}\sq{1}{6}\sq{2}{4}\sq{3}{5} + 15\ab{3}{5}\sq{1}{5}\sq{3}{4}\sq{3}{5} + 15\ab{5}{6}\sq{1}{4}\sq{3}{5}\sq{5}{6} \\
  & + 15\ab{5}{6}\sq{1}{6}\sq{3}{5}\sq{4}{5}
\end{align*}}
\noindent with $\mathcal{E}_{4}=N_{4}/(\ab{2}{4}^{2}\ab{3}{6}^{2}\sq{1}{4}^{2}\sq{2}{4}\sq{2}{5}\sq{3}{5}^{2})$.\par
\par\smallskip\noindent\textbf{Step 5}\nobreak
{\small
\begin{align*}
  N_{5} ={}& 3\ab{1}{4}\sq{1}{3}\sq{1}{4}\sq{4}{5} - 6\ab{1}{4}\sq{1}{4}^{2}\sq{3}{5} + 12\ab{1}{5}\sq{1}{4}\sq{1}{5}\sq{3}{5} \\
  & - 3\ab{1}{6}\sq{1}{4}\sq{1}{5}\sq{3}{6} - 12\ab{2}{3}\sq{1}{3}\sq{2}{4}\sq{3}{5} + 3\ab{2}{3}\sq{1}{5}\sq{2}{3}\sq{3}{4} \\
  & - 12\ab{2}{4}\sq{1}{4}\sq{2}{3}\sq{4}{5} - 12\ab{2}{4}\sq{1}{4}\sq{2}{5}\sq{3}{4} - 3\ab{2}{5}\underline{\sq{1}{3}}\sq{2}{5}\sq{4}{5} \\
  & + 3\ab{2}{5}\sq{1}{4}\sq{2}{5}\sq{3}{5} - 3\ab{2}{6}\sq{1}{5}\sq{2}{3}\sq{4}{6} - 12\ab{2}{6}\sq{1}{6}\sq{2}{4}\sq{3}{5} \\
  & + 15\ab{3}{5}\sq{1}{5}\sq{3}{4}\sq{3}{5} + 15\ab{5}{6}\sq{1}{4}\sq{3}{5}\sq{5}{6} + 15\ab{5}{6}\sq{1}{6}\sq{3}{5}\sq{4}{5}
\end{align*}}
\noindent with $\mathcal{E}_{5}=N_{5}/(\ab{2}{4}^{2}\ab{3}{6}^{2}\sq{1}{4}^{2}\sq{2}{4}\sq{2}{5}\sq{3}{5}^{2})$.\par
\par\smallskip\noindent\textbf{Step 6}\nobreak
{\small
\begin{align*}
  N_{6} ={}& 3\ab{1}{4}\sq{1}{3}\sq{1}{4}\sq{4}{5} - 6\ab{1}{4}\sq{1}{4}^{2}\sq{3}{5} + 12\ab{1}{5}\sq{1}{4}\sq{1}{5}\sq{3}{5} \\
  & - 3\ab{1}{6}\sq{1}{4}\sq{1}{5}\sq{3}{6} - 12\ab{2}{3}\sq{1}{3}\sq{2}{4}\sq{3}{5} + 3\ab{2}{3}\sq{1}{5}\sq{2}{3}\sq{3}{4} \\
  & - 12\ab{2}{4}\sq{1}{4}\sq{2}{3}\sq{4}{5} - 12\ab{2}{4}\sq{1}{4}\sq{2}{5}\sq{3}{4} + 3\ab{2}{5}\sq{1}{5}\sq{2}{5}\sq{3}{4} \\
  & - 3\ab{2}{6}\sq{1}{5}\sq{2}{3}\sq{4}{6} - 12\ab{2}{6}\sq{1}{6}\sq{2}{4}\sq{3}{5} + 15\ab{3}{5}\sq{1}{5}\sq{3}{4}\sq{3}{5} \\
  & + 15\ab{5}{6}\underline{\sq{1}{4}}\sq{3}{5}\sq{5}{6} + 15\ab{5}{6}\sq{1}{6}\sq{3}{5}\sq{4}{5}
\end{align*}}
\noindent with $\mathcal{E}_{6}=N_{6}/(\ab{2}{4}^{2}\ab{3}{6}^{2}\sq{1}{4}^{2}\sq{2}{4}\sq{2}{5}\sq{3}{5}^{2})$.\par
\par\smallskip\noindent\textbf{Step 7}\nobreak
{\small
\begin{align*}
  N_{7} ={}& 3\ab{1}{4}\sq{1}{3}\sq{1}{4}\sq{4}{5} - 6\ab{1}{4}\sq{1}{4}^{2}\sq{3}{5} + 12\ab{1}{5}\sq{1}{4}\sq{1}{5}\sq{3}{5} \\
  & - 3\ab{1}{6}\sq{1}{4}\sq{1}{5}\sq{3}{6} - 12\ab{2}{3}\sq{1}{3}\sq{2}{4}\sq{3}{5} + 3\ab{2}{3}\sq{1}{5}\sq{2}{3}\sq{3}{4} \\
  & - 12\ab{2}{4}\sq{1}{4}\sq{2}{3}\sq{4}{5} - 12\ab{2}{4}\sq{1}{4}\sq{2}{5}\sq{3}{4} + 3\ab{2}{5}\sq{1}{5}\sq{2}{5}\sq{3}{4} \\
  & - 3\ab{2}{6}\sq{1}{5}\sq{2}{3}\sq{4}{6} - 12\ab{2}{6}\sq{1}{6}\sq{2}{4}\sq{3}{5} + 15\underline{\ab{3}{5}}\sq{1}{5}\sq{3}{4}\sq{3}{5} \\
  & + 15\ab{5}{6}\sq{1}{5}\sq{3}{5}\sq{4}{6}
\end{align*}}
\noindent with $\mathcal{E}_{7}=N_{7}/(\ab{2}{4}^{2}\ab{3}{6}^{2}\sq{1}{4}^{2}\sq{2}{4}\sq{2}{5}\sq{3}{5}^{2})$.\par
\par\smallskip\noindent\textbf{Step 8}\nobreak
{\small
\begin{align*}
  N_{8} ={}& 3\ab{1}{4}\sq{1}{3}\sq{1}{4}\sq{4}{5} - 6\ab{1}{4}\sq{1}{4}^{2}\sq{3}{5} - 3\ab{1}{5}\sq{1}{4}\sq{1}{5}\sq{3}{5} \\
  & - 3\ab{1}{6}\sq{1}{4}\sq{1}{5}\sq{3}{6} - 12\underline{\ab{2}{3}}\sq{1}{3}\sq{2}{4}\sq{3}{5} + 3\ab{2}{3}\sq{1}{5}\sq{2}{3}\sq{3}{4} \\
  & - 12\ab{2}{4}\sq{1}{4}\sq{2}{3}\sq{4}{5} - 12\ab{2}{4}\sq{1}{4}\sq{2}{5}\sq{3}{4} - 15\ab{2}{5}\sq{1}{5}\sq{2}{4}\sq{3}{5} \\
  & + 3\ab{2}{5}\sq{1}{5}\sq{2}{5}\sq{3}{4} - 3\ab{2}{6}\sq{1}{5}\sq{2}{3}\sq{4}{6} - 12\ab{2}{6}\sq{1}{6}\sq{2}{4}\sq{3}{5}
\end{align*}}
\noindent with $\mathcal{E}_{8}=N_{8}/(\ab{2}{4}^{2}\ab{3}{6}^{2}\sq{1}{4}^{2}\sq{2}{4}\sq{2}{5}\sq{3}{5}^{2})$.\par
\par\smallskip\noindent\textbf{Step 9}\nobreak
{\small
\begin{align*}
  N_{9} ={}& 3\ab{1}{4}\sq{1}{3}\sq{1}{4}\sq{4}{5} - 6\ab{1}{4}\sq{1}{4}^{2}\sq{3}{5} - 3\ab{1}{5}\sq{1}{4}\sq{1}{5}\sq{3}{5} \\
  & - 3\ab{1}{6}\sq{1}{4}\sq{1}{5}\sq{3}{6} + 3\ab{2}{3}\sq{1}{5}\sq{2}{3}\sq{3}{4} - 12\ab{2}{4}\sq{1}{4}\sq{2}{3}\sq{4}{5} \\
  & + 12\ab{2}{4}\sq{1}{4}\underline{\sq{2}{4}}\sq{3}{5} - 12\ab{2}{4}\sq{1}{4}\sq{2}{5}\sq{3}{4} - 3\ab{2}{5}\sq{1}{5}\underline{\sq{2}{4}}\sq{3}{5} \\
  & + 3\ab{2}{5}\sq{1}{5}\sq{2}{5}\sq{3}{4} - 3\ab{2}{6}\sq{1}{5}\sq{2}{3}\sq{4}{6}
\end{align*}}
\noindent with $\mathcal{E}_{9}=N_{9}/(\ab{2}{4}^{2}\ab{3}{6}^{2}\sq{1}{4}^{2}\sq{2}{4}\sq{2}{5}\sq{3}{5}^{2})$.\par
\par\smallskip\noindent\textbf{Step 10}\nobreak
{\small
\begin{align*}
  N_{10} ={}& 3\ab{1}{4}\sq{1}{3}\sq{1}{4}\sq{4}{5} - 6\ab{1}{4}\sq{1}{4}^{2}\sq{3}{5} - 3\ab{1}{5}\sq{1}{4}\sq{1}{5}\sq{3}{5} \\
  & - 3\ab{1}{6}\sq{1}{4}\sq{1}{5}\sq{3}{6} + 3\underline{\ab{2}{3}}\sq{1}{5}\sq{2}{3}\sq{3}{4} - 3\ab{2}{5}\sq{1}{5}\sq{2}{3}\sq{4}{5} \\
  & - 3\ab{2}{6}\sq{1}{5}\sq{2}{3}\sq{4}{6}
\end{align*}}
\noindent with $\mathcal{E}_{10}=N_{10}/(\ab{2}{4}^{2}\ab{3}{6}^{2}\sq{1}{4}^{2}\sq{2}{4}\sq{2}{5}\sq{3}{5}^{2})$.\par
\par\smallskip\noindent\textbf{Step 11}\nobreak
{\small
\begin{align*}
  N_{11} ={}& 3\underline{\ab{1}{2}}\sq{1}{5}\sq{2}{3} + 3\ab{1}{4}\sq{1}{3}\sq{4}{5} - 6\ab{1}{4}\sq{1}{4}\sq{3}{5} \\
  & - 3\ab{1}{5}\sq{1}{5}\sq{3}{5} - 3\ab{1}{6}\sq{1}{5}\sq{3}{6}
\end{align*}}
\noindent with $\mathcal{E}_{11}=N_{11}/(\ab{2}{4}^{2}\ab{3}{6}^{2}\sq{1}{4}\sq{2}{4}\sq{2}{5}\sq{3}{5}^{2})$.\par
\par\smallskip\noindent\textbf{Step 12}\nobreak
\begin{equation*}\mathcal{E}_{12} = \frac{3\ab{1}{4}\underline{\sq{1}{3}}\sq{4}{5} - 6\ab{1}{4}\sq{1}{4}\sq{3}{5} + 3\ab{1}{4}\sq{1}{5}\sq{3}{4}}{\ab{2}{4}^{2}\ab{3}{6}^{2}\sq{1}{4}\sq{2}{4}\sq{2}{5}\sq{3}{5}^{2}}.\end{equation*}
\par\smallskip\noindent\textbf{Step 13}\nobreak
\begin{equation*}\mathcal{E}_{13} = \frac{-3\ab{1}{4}}{\ab{2}{4}^{2}\ab{3}{6}^{2}\sq{2}{4}\sq{2}{5}\sq{3}{5}}.\end{equation*}
}%
\newcommand{\exDeepCompareRows}{%
  \gbfs & yes & 1 & passed \\
  SFT & no & 102 & passed \\
  CDS ($B{=}20$) & no & 4 & failed \\
}%
\section{Example derivations}
\label{app:examples}

This appendix gives explicit derivations for the sum-denominator input of Fig.~\ref{fig:worked} and for a six-point case from the size$\times$depth suite. For each derivation, a table lists the identity applied at each step and the resulting bracket and term counts, and the intermediate expressions follow the table.

\subsection{Cancellation of a Dirac-chain factor}
\label{app:sumden-example}
The synthetic input illustrated in Fig.~\ref{fig:worked} is
\exSumdenInput
\noindent with
\exSumdenNumerator
\noindent \gbfs\ reaches $\ab{3}{4}$ after evaluating $2{,}672$ moves.
\par\begin{center}\footnotesize\setlength{\tabcolsep}{4pt}\renewcommand{\arraystretch}{1.35}
\begin{threeparttable}
\begin{tabular}{@{}clcrrr@{}}
  \toprule
  & & & \multicolumn{2}{c}{brackets} & \\
  \cmidrule(lr){4-5}
  step & identity & rewrite & raw & factored & terms \\
  \midrule
\exSumdenStepRows
  \bottomrule
\end{tabular}
\par\vspace{2pt}\noindent\parbox{\linewidth}{\raggedright \hypertarget{app:sum-denominator-star}{$^{*}$}\,Starred steps replace one bracket factor in a single numerator term. Unstarred steps replace every occurrence of the indicated bracket in the numerator. Underlined brackets are replaced in the next step.}
\end{threeparttable}
\end{center}
\exSumdenWideRules
\exSumdenSteps

\subsection{A six-point case from the size$\times$depth suite}
\label{app:case144}
We consider six-point case $144$ from the size$\times$depth suite,
with generation depth $K=30$ and bracket limit $L=120$.
Of the $30$ scrambling moves, $5$ multiply a term by a unity factor
and $6$ add a vanishing combination.
The input has $23$ terms and a raw numerator bracket count of $115$,
whereas the target has a single numerator bracket.
\gbfs\ recovers the target in a $13$-move derivation,
found after evaluating $31{,}108$ moves.
Table~\ref{tab:case144} compares the three methods on this input.
\par\begin{center}\footnotesize\setlength{\tabcolsep}{4pt}\renewcommand{\arraystretch}{1.35}
\begin{threeparttable}
\begin{tabular}{@{}clcrrr@{}}
  \toprule
  & & & \multicolumn{2}{c}{brackets} & \\
  \cmidrule(lr){4-5}
  step & identity & rewrite & raw & factored & terms \\
  \midrule
\exDeepStepRows
  \bottomrule
\end{tabular}
\par\vspace{2pt}\noindent\parbox{\linewidth}{\raggedright \hypertarget{app:case144-star}{$^{*}$}\,Starred steps replace one bracket factor in a single numerator term. Unstarred steps replace every occurrence of the indicated bracket in the numerator. Underlined brackets are replaced in the next step.}
\end{threeparttable}
\end{center}
\exDeepWideRules
\noindent The target is
\exDeepTarget
\exDeepSteps
\begin{table}[htbp]
\centering\small\renewcommand{\gbfs}{\textbf{GBFS}}\renewcommand{\arraystretch}{1.25}
\begin{tabular}{@{}lcrc@{}}
  \toprule
  method & solved & raw brackets & finite-field check \\
  \midrule
\exDeepCompareRows
  \bottomrule
\end{tabular}
\caption{Comparison of \gbfs, SFT, and CDS on six-point case $144$ with $K=30$ and $L=120$. The target has a single numerator bracket. The last column gives the result of the finite-field check of each output against the input. CDS uses the inference settings described in \S\ref{sec:public}, and its row shows the highest-scoring of the $20$ beam candidates. None of these candidates passes the finite-field check.}
\label{tab:case144}
\end{table}

\FloatBarrier
\providecommand{\bcfit}[1]{\resizebox{\ifdim\width>\linewidth\linewidth\else\width\fi}{!}{$\displaystyle #1$}}%
\newcommand{\belowListing}{%
\par\medskip\noindent\textbf{case 9042} \quad $21 \to 7$ brackets ($\Delta=-14$):\par\nobreak
\noindent\bcfit{\text{target:}\;\; \frac{- \sq{1}{3}\sq{2}{4}^{2}\sq{2}{5}^{3}\sq{4}{5} + \sq{1}{4}\sq{2}{4}^{2}\sq{2}{5}^{3}\sq{3}{5} + \sq{1}{5}\sq{2}{3}\sq{2}{4}^{2}\sq{2}{5}^{2}\sq{4}{5}}{\ab{1}{2}^{2}\ab{1}{4}^{2}\ab{2}{5}\ab{3}{5}}}\par\nobreak\vspace{3pt}
\noindent\bcfit{\text{\gbfs:}\;\; \frac{\sq{1}{5}\sq{2}{4}^{3}\sq{2}{5}^{2}\sq{3}{5}}{\ab{1}{2}^{2}\ab{1}{4}^{2}\ab{2}{5}\ab{3}{5}}}\par
\par\medskip\noindent\textbf{case 288} \quad $12 \to 0$ brackets ($\Delta=-12$):\par\nobreak
\noindent\bcfit{\text{target:}\;\; \frac{\ab{2}{5}\ab{4}{5}\sq{1}{2}\sq{2}{5} + \ab{3}{5}\ab{4}{5}\sq{1}{3}\sq{2}{5} + \ab{4}{5}^{2}\sq{1}{4}\sq{2}{5}}{\sq{2}{4}}}\par\nobreak\vspace{3pt}
\noindent\bcfit{\text{\gbfs:}\;\; 0}\par
\par\medskip\noindent\textbf{case 4450} \quad $30 \to 20$ brackets ($\Delta=-10$):\par\nobreak
\noindent\bcfit{\text{target:}\;\; \frac{\ab{1}{2}\ab{1}{4}\ab{2}{3}\ab{3}{4}^{2}\ab{3}{5}\ab{4}{5}\sq{1}{2}^{2}\sq{2}{5} - \ab{1}{3}\ab{1}{4}\ab{2}{3}\ab{2}{4}\ab{3}{4}\ab{3}{5}\ab{4}{5}\sq{1}{2}^{2}\sq{2}{5} + \ab{1}{4}\ab{2}{3}\ab{2}{5}\ab{3}{4}^{3}\ab{3}{5}\sq{1}{2}\sq{2}{3}\sq{2}{5}}{\ab{2}{4}\sq{1}{5}}}\par\nobreak\vspace{3pt}
\noindent\bcfit{\text{\gbfs:}\;\; \frac{\ab{1}{4}\ab{2}{3}\ab{2}{5}\ab{3}{4}^{3}\ab{3}{5}\sq{1}{2}\sq{2}{3}\sq{2}{5} - \ab{1}{4}^{2}\ab{2}{3}^{2}\ab{3}{4}\ab{3}{5}\ab{4}{5}\sq{1}{2}^{2}\sq{2}{5}}{\ab{2}{4}\sq{1}{5}}}\par
\par\medskip\noindent\textbf{case 7144} \quad $27 \to 18$ brackets ($\Delta=-9$):\par\nobreak
\noindent\bcfit{\text{target:}\;\; - \ab{1}{4}\ab{2}{5}\ab{3}{4}\ab{4}{5}\sq{1}{5}^{3}\sq{2}{3}\sq{3}{5} - \ab{2}{4}^{3}\ab{3}{4}\sq{1}{4}\sq{1}{5}\sq{2}{3}^{2}\sq{2}{5} - \ab{2}{4}^{2}\ab{3}{4}^{2}\sq{1}{4}\sq{1}{5}\sq{2}{3}^{2}\sq{3}{5}}\par\nobreak\vspace{3pt}
\noindent\bcfit{\text{\gbfs:}\;\; - \ab{1}{4}\ab{2}{3}\ab{2}{4}\ab{3}{4}\sq{1}{3}\sq{1}{5}^{2}\sq{2}{3}^{2} + \ab{1}{4}^{2}\ab{2}{5}\ab{3}{4}\sq{1}{3}\sq{1}{5}^{3}\sq{2}{3}}\par
\par\medskip\noindent\textbf{case 1930} \quad $24 \to 16$ brackets ($\Delta=-8$):\par\nobreak
\noindent\bcfit{\text{target:}\;\; \frac{- \ab{1}{2}^{2}\ab{2}{3}\ab{2}{4}\ab{2}{5}^{3}\sq{3}{4} + \ab{1}{2}^{2}\ab{2}{3}\ab{2}{5}^{4}\sq{3}{5} + \ab{1}{2}^{2}\ab{2}{4}\ab{2}{5}^{4}\sq{4}{5}}{\ab{2}{4}\ab{4}{5}^{2}\sq{1}{4}\sq{1}{5}}}\par\nobreak\vspace{3pt}
\noindent\bcfit{\text{\gbfs:}\;\; \frac{\ab{1}{2}^{2}\ab{2}{3}\ab{2}{5}^{4}\sq{3}{5} - \ab{1}{2}^{3}\ab{2}{4}\ab{2}{5}^{3}\sq{1}{4}}{\ab{2}{4}\ab{4}{5}^{2}\sq{1}{4}\sq{1}{5}}}\par
\par\medskip\noindent\textbf{case 1134} \quad $12 \to 5$ brackets ($\Delta=-7$):\par\nobreak
\noindent\bcfit{\text{target:}\;\; \frac{\ab{1}{2}\ab{1}{3}^{2}\ab{2}{3}\sq{2}{4}\sq{2}{5} + \ab{1}{3}^{3}\ab{2}{3}\sq{2}{4}\sq{3}{5}}{\ab{2}{5}\sq{2}{5}\sq{4}{5}}}\par\nobreak\vspace{3pt}
\noindent\bcfit{\text{\gbfs:}\;\; \frac{- \ab{1}{3}^{2}\ab{1}{4}\ab{2}{3}\sq{2}{4}}{\ab{2}{5}\sq{2}{5}}}\par
\par\medskip\noindent\textbf{case 2282} \quad $12 \to 5$ brackets ($\Delta=-7$):\par\nobreak
\noindent\bcfit{\text{target:}\;\; \frac{\ab{1}{3}\ab{2}{4}^{2}\sq{1}{5}^{2}\sq{2}{5} + \ab{2}{3}\ab{2}{4}^{2}\sq{1}{5}\sq{2}{5}^{2}}{\ab{1}{4}\ab{1}{5}\ab{4}{5}^{3}\sq{1}{2}^{3}\sq{2}{4}^{4}\sq{3}{5}^{2}\sq{4}{5}}}\par\nobreak\vspace{3pt}
\noindent\bcfit{\text{\gbfs:}\;\; \frac{\ab{2}{4}^{2}\ab{3}{4}\sq{1}{5}\sq{2}{5}}{\ab{1}{4}\ab{1}{5}\ab{4}{5}^{3}\sq{1}{2}^{3}\sq{2}{4}^{4}\sq{3}{5}^{2}}}\par
\par\medskip\noindent\textbf{case 6970} \quad $21 \to 14$ brackets ($\Delta=-7$):\par\nobreak
\noindent\bcfit{\text{target:}\;\; \frac{\ab{2}{3}\ab{2}{4}\ab{4}{5}\sq{1}{2}^{2}\sq{1}{3}^{2} + \ab{2}{4}^{2}\ab{2}{5}\sq{1}{2}^{3}\sq{1}{3} + \ab{2}{4}^{2}\ab{3}{5}\sq{1}{2}^{2}\sq{1}{3}^{2}}{\sq{2}{3}}}\par\nobreak\vspace{3pt}
\noindent\bcfit{\text{\gbfs:}\;\; \frac{\ab{2}{3}\ab{2}{4}\ab{4}{5}\sq{1}{2}^{2}\sq{1}{3}^{2} - \ab{2}{4}^{2}\ab{4}{5}\sq{1}{2}^{2}\sq{1}{3}\sq{1}{4}}{\sq{2}{3}}}\par
\par\medskip\noindent\textbf{case 814} \quad $9 \to 3$ brackets ($\Delta=-6$):\par\nobreak
\noindent\bcfit{\text{target:}\;\; \frac{\sq{1}{2}\sq{2}{4}\sq{3}{4} + \sq{1}{3}\sq{2}{4}^{2} + \sq{1}{4}\sq{2}{3}\sq{2}{4}}{\ab{4}{5}\sq{1}{4}\sq{2}{3}}}\par\nobreak\vspace{3pt}
\noindent\bcfit{\text{\gbfs:}\;\; \frac{2\sq{1}{3}\sq{2}{4}^{2}}{\ab{4}{5}\sq{1}{4}\sq{2}{3}}}\par
\par\medskip\noindent\textbf{case 1468} \quad $18 \to 12$ brackets ($\Delta=-6$):\par\nobreak
\noindent\bcfit{\text{target:}\;\; \frac{- \ab{1}{3}\ab{1}{4}\ab{2}{3}^{2}\sq{1}{3}\sq{4}{5} + \ab{1}{3}\ab{2}{3}^{2}\ab{2}{4}\sq{2}{3}\sq{4}{5} + \ab{1}{5}\ab{2}{3}^{2}\ab{2}{4}\sq{2}{5}\sq{4}{5}}{\ab{2}{5}\ab{4}{5}\sq{1}{4}}}\par\nobreak\vspace{3pt}
\noindent\bcfit{\text{\gbfs:}\;\; \frac{- \ab{1}{3}\ab{1}{4}\ab{2}{3}^{2}\sq{1}{3}\sq{4}{5} - \ab{1}{4}\ab{2}{3}^{2}\ab{2}{4}\sq{2}{4}\sq{4}{5}}{\ab{2}{5}\ab{4}{5}\sq{1}{4}}}\par
\par\medskip\noindent\textbf{case 3732} \quad $12 \to 6$ brackets ($\Delta=-6$):\par\nobreak
\noindent\bcfit{\text{target:}\;\; \frac{\ab{1}{2}\ab{1}{4}\sq{1}{2}\sq{2}{3}^{2}\sq{3}{4} - \ab{1}{2}\ab{3}{4}\sq{2}{3}^{3}\sq{3}{4}}{\sq{1}{3}\sq{3}{5}^{2}}}\par\nobreak\vspace{3pt}
\noindent\bcfit{\text{\gbfs:}\;\; \frac{- \ab{1}{2}\ab{4}{5}\sq{2}{3}^{2}\sq{2}{5}\sq{3}{4}}{\sq{1}{3}\sq{3}{5}^{2}}}\par
\par\medskip\noindent\textbf{case 6124} \quad $9 \to 3$ brackets ($\Delta=-6$):\par\nobreak
\noindent\bcfit{\text{target:}\;\; \frac{\ab{1}{5}\ab{3}{4}\sq{1}{2} + \ab{3}{4}\ab{3}{5}\sq{2}{3} - \ab{3}{4}\ab{4}{5}\sq{2}{4}}{\ab{1}{2}\ab{2}{3}\ab{4}{5}\sq{1}{2}\sq{1}{4}^{2}\sq{1}{5}}}\par\nobreak\vspace{3pt}
\noindent\bcfit{\text{\gbfs:}\;\; \frac{2\ab{3}{4}\ab{3}{5}\sq{2}{3}}{\ab{1}{2}\ab{2}{3}\ab{4}{5}\sq{1}{2}\sq{1}{4}^{2}\sq{1}{5}}}\par
\par\medskip\noindent\textbf{case 7034} \quad $18 \to 12$ brackets ($\Delta=-6$):\par\nobreak
\noindent\bcfit{\text{target:}\;\; \frac{- \ab{1}{2}\ab{1}{3}\ab{1}{4}\ab{2}{5}\sq{2}{4}\sq{4}{5} - \ab{1}{3}^{3}\ab{2}{4}\sq{3}{4}^{2} - \ab{1}{3}^{3}\ab{2}{5}\sq{3}{4}\sq{3}{5}}{\sq{1}{4}}}\par\nobreak\vspace{3pt}
\noindent\bcfit{\text{\gbfs:}\;\; \frac{- \ab{1}{2}\ab{1}{3}\ab{1}{4}\ab{2}{5}\sq{2}{4}\sq{4}{5} + \ab{1}{2}\ab{1}{3}^{3}\sq{1}{3}\sq{3}{4}}{\sq{1}{4}}}\par
\par\medskip\noindent\textbf{case 7624} \quad $9 \to 3$ brackets ($\Delta=-6$):\par\nobreak
\noindent\bcfit{\text{target:}\;\; \frac{\sq{1}{2}^{2}\sq{3}{5} + \sq{1}{2}\sq{1}{3}\sq{2}{5} - \sq{1}{2}\sq{1}{5}\sq{2}{3}}{\ab{2}{5}^{2}\sq{2}{5}\sq{3}{4}}}\par\nobreak\vspace{3pt}
\noindent\bcfit{\text{\gbfs:}\;\; \frac{2\sq{1}{2}^{2}\sq{3}{5}}{\ab{2}{5}^{2}\sq{2}{5}\sq{3}{4}}}\par
\par\medskip\noindent\textbf{case 750} \quad $10 \to 5$ brackets ($\Delta=-5$):\par\nobreak
\noindent\bcfit{\text{target:}\;\; \frac{\ab{1}{3}\sq{2}{3}\sq{3}{5}^{2}\sq{4}{5} + \ab{1}{4}\sq{2}{4}\sq{3}{5}^{2}\sq{4}{5}}{\ab{4}{5}^{2}\sq{1}{2}^{2}\sq{1}{4}^{2}\sq{3}{4}^{2}}}\par\nobreak\vspace{3pt}
\noindent\bcfit{\text{\gbfs:}\;\; \frac{- \ab{1}{5}\sq{2}{5}\sq{3}{5}^{2}\sq{4}{5}}{\ab{4}{5}^{2}\sq{1}{2}^{2}\sq{1}{4}^{2}\sq{3}{4}^{2}}}\par
\par\medskip\noindent\textbf{case 1102} \quad $15 \to 10$ brackets ($\Delta=-5$):\par\nobreak
\noindent\bcfit{\text{target:}\;\; \frac{\ab{1}{2}\ab{1}{4}\ab{1}{5}\ab{2}{3}\ab{3}{4} + \ab{1}{3}\ab{1}{4}^{2}\ab{2}{3}\ab{2}{5} - \ab{1}{4}^{2}\ab{1}{5}\ab{2}{3}^{2}}{\ab{3}{4}\ab{3}{5}\sq{1}{3}\sq{1}{5}\sq{2}{3}\sq{2}{4}}}\par\nobreak\vspace{3pt}
\noindent\bcfit{\text{\gbfs:}\;\; \frac{\ab{1}{2}\ab{1}{4}\ab{1}{5}\ab{2}{3}\ab{3}{4} + \ab{1}{2}\ab{1}{4}^{2}\ab{2}{3}\ab{3}{5}}{\ab{3}{4}\ab{3}{5}\sq{1}{3}\sq{1}{5}\sq{2}{3}\sq{2}{4}}}\par
\par\medskip\noindent\textbf{case 2584} \quad $15 \to 10$ brackets ($\Delta=-5$):\par\nobreak
\noindent\bcfit{\text{target:}\;\; \frac{- \ab{1}{2}\ab{4}{5}\sq{1}{3}\sq{1}{4}\sq{3}{4} + \ab{2}{4}\ab{4}{5}\sq{1}{4}\sq{3}{4}^{2} + \ab{2}{5}^{2}\sq{1}{4}\sq{2}{3}\sq{3}{5}}{\ab{1}{3}\ab{2}{3}\ab{2}{5}^{2}\ab{3}{5}\sq{1}{2}\sq{1}{5}\sq{2}{4}\sq{4}{5}^{2}}}\par\nobreak\vspace{3pt}
\noindent\bcfit{\text{\gbfs:}\;\; \frac{- 2\ab{1}{2}\ab{4}{5}\sq{1}{3}\sq{1}{4}\sq{3}{4} - \ab{1}{5}\ab{2}{5}\sq{1}{3}\sq{1}{4}\sq{3}{5}}{\ab{1}{3}\ab{2}{3}\ab{2}{5}^{2}\ab{3}{5}\sq{1}{2}\sq{1}{5}\sq{2}{4}\sq{4}{5}^{2}}}\par
\par\medskip\noindent\textbf{case 7272} \quad $15 \to 10$ brackets ($\Delta=-5$):\par\nobreak
\noindent\bcfit{\text{target:}\;\; \frac{\ab{1}{2}\ab{3}{5}\sq{1}{5}\sq{3}{5}\sq{4}{5} - \ab{1}{3}\ab{2}{5}\sq{1}{5}\sq{3}{5}\sq{4}{5} + \ab{2}{3}\ab{4}{5}\sq{3}{5}\sq{4}{5}^{2}}{\ab{2}{4}^{3}\ab{3}{4}\sq{1}{3}\sq{3}{4}^{3}}}\par\nobreak\vspace{3pt}
\noindent\bcfit{\text{\gbfs:}\;\; \frac{- \ab{1}{5}\ab{2}{3}\sq{1}{5}\sq{3}{5}\sq{4}{5} + \ab{2}{3}\ab{4}{5}\sq{3}{5}\sq{4}{5}^{2}}{\ab{2}{4}^{3}\ab{3}{4}\sq{1}{3}\sq{3}{4}^{3}}}\par
\par\medskip\noindent\textbf{case 7560} \quad $15 \to 10$ brackets ($\Delta=-5$):\par\nobreak
\noindent\bcfit{\text{target:}\;\; \frac{- \ab{1}{2}\ab{1}{5}^{2}\sq{1}{4}\sq{2}{4} + \ab{1}{3}\ab{1}{5}^{2}\sq{1}{4}\sq{3}{4} - \ab{1}{5}^{2}\ab{2}{5}\sq{2}{4}\sq{4}{5}}{\ab{2}{3}\sq{1}{3}\sq{2}{3}}}\par\nobreak\vspace{3pt}
\noindent\bcfit{\text{\gbfs:}\;\; \frac{\ab{1}{3}\ab{1}{5}^{2}\sq{1}{4}\sq{3}{4} - \ab{1}{5}^{2}\ab{2}{3}\sq{2}{4}\sq{3}{4}}{\ab{2}{3}\sq{1}{3}\sq{2}{3}}}\par
\par\medskip\noindent\textbf{case 3988} \quad $12 \to 8$ brackets ($\Delta=-4$):\par\nobreak
\noindent\bcfit{\text{target:}\;\; \frac{\ab{3}{4}^{2}\ab{3}{5}\sq{1}{3} - \ab{3}{4}^{2}\ab{4}{5}\sq{1}{4} - \ab{3}{4}\ab{3}{5}\ab{4}{5}\sq{1}{5}}{\sq{4}{5}}}\par\nobreak\vspace{3pt}
\noindent\bcfit{\text{\gbfs:}\;\; \frac{- \ab{2}{4}\ab{3}{4}\ab{3}{5}\sq{1}{2} - \ab{3}{4}^{2}\ab{4}{5}\sq{1}{4}}{\sq{4}{5}}}\par
\par\medskip\noindent\textbf{case 4706} \quad $4 \to 0$ brackets ($\Delta=-4$):\par\nobreak
\noindent\bcfit{\text{target:}\;\; \frac{- \sq{1}{4}\sq{2}{5} + \sq{1}{5}\sq{2}{4}}{\ab{2}{3}\sq{1}{2}\sq{1}{4}\sq{3}{4}^{2}\sq{4}{5}}}\par\nobreak\vspace{3pt}
\noindent\bcfit{\text{\gbfs:}\;\; \frac{- 1}{\ab{2}{3}\sq{1}{4}\sq{3}{4}^{2}}}\par
\par\medskip\noindent\textbf{case 4976} \quad $12 \to 8$ brackets ($\Delta=-4$):\par\nobreak
\noindent\bcfit{\text{target:}\;\; \frac{\ab{1}{2}^{2}\ab{2}{5}\sq{2}{4} - \ab{1}{2}^{2}\ab{3}{5}\sq{3}{4} + \ab{1}{2}\ab{1}{3}\ab{2}{5}\sq{3}{4}}{\ab{1}{5}\ab{2}{3}^{3}\sq{2}{4}^{2}}}\par\nobreak\vspace{3pt}
\noindent\bcfit{\text{\gbfs:}\;\; \frac{\ab{1}{2}\ab{1}{5}\ab{2}{5}\sq{4}{5} - \ab{1}{2}^{2}\ab{3}{5}\sq{3}{4}}{\ab{1}{5}\ab{2}{3}^{3}\sq{2}{4}^{2}}}\par
\par\medskip\noindent\textbf{case 5470} \quad $12 \to 8$ brackets ($\Delta=-4$):\par\nobreak
\noindent\bcfit{\text{target:}\;\; \frac{\ab{1}{2}\ab{1}{4}\sq{1}{2}^{2} + \ab{1}{4}^{2}\sq{1}{2}\sq{1}{4} - \ab{1}{5}\ab{2}{4}\sq{1}{2}\sq{2}{5}}{\ab{1}{3}^{2}\ab{4}{5}\sq{2}{4}\sq{2}{5}\sq{3}{5}}}\par\nobreak\vspace{3pt}
\noindent\bcfit{\text{\gbfs:}\;\; \frac{- \ab{1}{3}\ab{1}{4}\sq{1}{2}\sq{1}{3} + \ab{1}{5}\ab{3}{4}\sq{1}{2}\sq{3}{5}}{\ab{1}{3}^{2}\ab{4}{5}\sq{2}{4}\sq{2}{5}\sq{3}{5}}}\par
\par\medskip\noindent\textbf{case 6444} \quad $6 \to 2$ brackets ($\Delta=-4$):\par\nobreak
\noindent\bcfit{\text{target:}\;\; \frac{- \ab{1}{2}^{2}\sq{1}{4} - \ab{1}{2}\ab{2}{5}\sq{4}{5}}{\sq{3}{4}}}\par\nobreak\vspace{3pt}
\noindent\bcfit{\text{\gbfs:}\;\; - \ab{1}{2}\ab{2}{3}}\par
\par\medskip\noindent\textbf{case 7240} \quad $12 \to 8$ brackets ($\Delta=-4$):\par\nobreak
\noindent\bcfit{\text{target:}\;\; \frac{- \ab{1}{2}^{2}\ab{1}{3}\sq{1}{4} - \ab{1}{2}\ab{1}{3}\ab{2}{3}\sq{3}{4} + \ab{1}{2}\ab{1}{5}\ab{2}{3}\sq{4}{5}}{\ab{2}{5}\ab{3}{4}\sq{1}{5}\sq{2}{4}^{2}\sq{2}{5}\sq{3}{5}}}\par\nobreak\vspace{3pt}
\noindent\bcfit{\text{\gbfs:}\;\; \frac{- \ab{1}{2}^{2}\ab{1}{3}\sq{1}{4} + \ab{1}{2}^{2}\ab{2}{3}\sq{2}{4}}{\ab{2}{5}\ab{3}{4}\sq{1}{5}\sq{2}{4}^{2}\sq{2}{5}\sq{3}{5}}}\par
\par\medskip\noindent\textbf{case 7464} \quad $6 \to 2$ brackets ($\Delta=-4$):\par\nobreak
\noindent\bcfit{\text{target:}\;\; \frac{\ab{1}{3}\sq{2}{3}\sq{2}{5} + \ab{1}{5}\sq{2}{5}^{2}}{\ab{2}{3}\sq{1}{2}\sq{2}{4}^{2}}}\par\nobreak\vspace{3pt}
\noindent\bcfit{\text{\gbfs:}\;\; \frac{- \ab{1}{4}\sq{2}{5}}{\ab{2}{3}\sq{1}{2}\sq{2}{4}}}\par
\par\medskip\noindent\textbf{case 878} \quad $6 \to 3$ brackets ($\Delta=-3$):\par\nobreak
\noindent\bcfit{\text{target:}\;\; - \ab{1}{2}\sq{2}{5}^{2} - \ab{1}{4}\sq{2}{5}\sq{4}{5}}\par\nobreak\vspace{3pt}
\noindent\bcfit{\text{\gbfs:}\;\; \ab{1}{3}\sq{2}{5}\sq{3}{5}}\par
\par\medskip\noindent\textbf{case 1372} \quad $6 \to 3$ brackets ($\Delta=-3$):\par\nobreak
\noindent\bcfit{\text{target:}\;\; \frac{- \ab{1}{3}\ab{3}{4}\sq{3}{5} - \ab{1}{4}\ab{3}{4}\sq{4}{5}}{\ab{2}{4}\sq{1}{3}^{2}\sq{2}{3}\sq{2}{4}}}\par\nobreak\vspace{3pt}
\noindent\bcfit{\text{\gbfs:}\;\; \frac{\ab{1}{2}\ab{3}{4}\sq{2}{5}}{\ab{2}{4}\sq{1}{3}^{2}\sq{2}{3}\sq{2}{4}}}\par
\par\medskip\noindent\textbf{case 6220} \quad $6 \to 3$ brackets ($\Delta=-3$):\par\nobreak
\noindent\bcfit{\text{target:}\;\; \frac{- \ab{1}{3}\ab{3}{5}\sq{1}{2} - \ab{3}{4}\ab{3}{5}\sq{2}{4}}{\ab{1}{4}\sq{1}{5}}}\par\nobreak\vspace{3pt}
\noindent\bcfit{\text{\gbfs:}\;\; \frac{\ab{3}{5}^{2}\sq{2}{5}}{\ab{1}{4}\sq{1}{5}}}\par
\par\medskip\noindent\textbf{case 7798} \quad $6 \to 3$ brackets ($\Delta=-3$):\par\nobreak
\noindent\bcfit{\text{target:}\;\; \frac{- \ab{1}{2}^{2}\ab{3}{4} - \ab{1}{2}\ab{1}{4}\ab{2}{3}}{\sq{3}{4}}}\par\nobreak\vspace{3pt}
\noindent\bcfit{\text{\gbfs:}\;\; \frac{- \ab{1}{2}\ab{1}{3}\ab{2}{4}}{\sq{3}{4}}}\par
\par\medskip\noindent\textbf{case 8850} \quad $9 \to 6$ brackets ($\Delta=-3$):\par\nobreak
\noindent\bcfit{\text{target:}\;\; \frac{\ab{1}{3}\ab{2}{5}\sq{1}{3} - \ab{1}{5}\ab{2}{3}\sq{1}{3} + \ab{1}{5}\ab{2}{5}\sq{1}{5}}{\ab{1}{4}\ab{2}{4}\sq{1}{2}\sq{3}{4}^{3}}}\par\nobreak\vspace{3pt}
\noindent\bcfit{\text{\gbfs:}\;\; \frac{\ab{1}{2}\ab{3}{5}\sq{1}{3} + \ab{1}{5}\ab{2}{5}\sq{1}{5}}{\ab{1}{4}\ab{2}{4}\sq{1}{2}\sq{3}{4}^{3}}}\par
\par\medskip\noindent\textbf{case 1166} \quad $20 \to 18$ brackets ($\Delta=-2$):\par\nobreak
\noindent\bcfit{\text{target:}\;\; \frac{- \ab{1}{2}\ab{1}{4}\ab{2}{3}\ab{2}{4}\ab{3}{4}\ab{3}{5}\ab{4}{5}\sq{1}{3}\sq{1}{4}\sq{1}{5} + \ab{1}{4}\ab{1}{5}\ab{2}{3}^{3}\ab{3}{4}\ab{4}{5}\sq{1}{3}^{2}\sq{1}{5}}{\ab{2}{5}^{2}}}\par\nobreak\vspace{3pt}
\noindent\bcfit{\text{\gbfs:}\;\; \frac{- \ab{1}{3}\ab{1}{4}\ab{2}{3}\ab{2}{4}\ab{3}{4}\ab{4}{5}\sq{1}{3}\sq{1}{4}\sq{1}{5} - \ab{1}{4}\ab{1}{5}\ab{2}{3}^{2}\ab{3}{4}\ab{4}{5}\sq{1}{3}\sq{1}{5}^{2}}{\ab{2}{5}}}\par
}%
\section{Cases simplified below the target}
\label{app:below}

The target of a case is not a proven minimum. In target-free mode, we ran \gbfs\ on a sample of $296$ cases labeled as non-vanishing in the five-point public set, with a budget of $3\times10^{4}$ moves per case. No case reached the $60\,\mathrm{s}$ time cap. For $264$ cases, the output has the same raw numerator bracket count as the target. For the other $32$, it has $2$ to $14$ fewer brackets.

In $25$ of these $32$ cases, the output has the same denominator as the target. In $6$ cases, the denominator of the output is a proper divisor of that of the target. In the remaining case $288$, the target vanishes by momentum conservation and the output is $0$.

The following list gives the target and search output for each case. Case numbers
refer to the subset labeled as non-vanishing in the five-point public set. Each heading gives the
raw numerator bracket counts of the target and the output, together with the change $\Delta$ from
the target to the output. Cases are ordered by $\Delta$. In case
$4706$, the output has a nonzero constant numerator and therefore a bracket count of $0$.

\begingroup\footnotesize
\belowListing
\endgroup

\clearpage
\section{Full results on the size$\times$depth suite}
\label{app:suite}
Table~\ref{tab:capdepthext} gives the solve rates on all $72$ cells of the size$\times$depth suite, and Table~\ref{tab:budgetapps} gives the number of moves used per solved case.

\begin{table}[!htbp]
  \centering
  \footnotesize
  \setlength{\tabcolsep}{3pt}
  {\resizebox{\textwidth}{!}{%
  \begin{tabular}{llcccccccccccc}
    \toprule
    & & \multicolumn{2}{c}{$L=60$} & \multicolumn{2}{c}{$L=90$} & \multicolumn{2}{c}{$L=120$} & \multicolumn{2}{c}{$L=150$} & \multicolumn{2}{c}{$L=200$} & \multicolumn{2}{c}{$L=250$} \\
    \cmidrule(lr){3-4}\cmidrule(lr){5-6}\cmidrule(lr){7-8}\cmidrule(lr){9-10}\cmidrule(lr){11-12}\cmidrule(lr){13-14}
    $n$ & $K$ & \textbf{GBFS}/$+$ & SFT & \textbf{GBFS}/$+$ & SFT & \textbf{GBFS}/$+$ & SFT & \textbf{GBFS}/$+$ & SFT & \textbf{GBFS}/$+$ & SFT & \textbf{GBFS}/$+$ & SFT \\
    \midrule
    \multirow{4}{*}{$4$} & $15$ & $\mathbf{100}$ & $98.33$ & $\mathbf{100}$ & $97.33$ & $\mathbf{100}$ & $96.67^{\dagger}$ & $99.67$/$\mathbf{100}$ & $89.67^{\dagger}$ & $99.67$/$\mathbf{100}$ & $79.33^{\dagger}$ & $99.67$/$\mathbf{100}$ & $72.33^{\dagger}$ \\
    & $20$ & $\mathbf{100}$ & $99.00$ & $\mathbf{100}$ & $98.67$ & $99.67$/$\mathbf{100}$ & $94.33^{\dagger}$ & $99.67$/$\mathbf{100}$ & $88.67^{\dagger}$ & $99.67$/$\mathbf{100}$ & $79.00^{\dagger}$ & $99.67$/$\mathbf{100}$ & $63.33^{\dagger}$ \\
    & $25$ & $\mathbf{100}$ & $98.67$ & $\mathbf{100}$ & $96.67$ & $99.33$/$\mathbf{100}$ & $96.00^{\dagger}$ & $99.00$/$99.67^{\ddagger}$ & $85.00^{\dagger}$ & $99.67$/$\mathbf{100}$ & $77.00^{\dagger}$ & $99.33$/$\mathbf{100}$ & $64.00^{\dagger}$ \\
    & $30$ & $\mathbf{100}$ & $98.33$ & $99.67$/$\mathbf{100}$ & $98.00$ & $\mathbf{100}$ & $92.67^{\dagger}$ & $\mathbf{100}$ & $90.00^{\dagger}$ & $\mathbf{100}$ & $75.33^{\dagger}$ & $\mathbf{100}$ & $59.67^{\dagger}$ \\
    \midrule
    \multirow{4}{*}{$5$} & $15$ & $\mathbf{100}$ & $\mathbf{100}$ & $\mathbf{100}$ & $99.67$ & $\mathbf{100}$ & $96.67$ & $\mathbf{100}$ & $88.67$ & $\mathbf{100}$ & $84.67$ & $99.33$/$\mathbf{100}$ & $71.00$ \\
    & $20$ & $\mathbf{100}$ & $99.33$ & $\mathbf{100}$ & $99.00$ & $\mathbf{100}$ & $95.67$ & $\mathbf{100}$ & $90.00$ & $\mathbf{100}$ & $77.33$ & $99.00$/$\mathbf{100}$ & $62.00$ \\
    & $25$ & $\mathbf{100}$ & $99.67$ & $\mathbf{100}$ & $97.67$ & $\mathbf{100}$ & $96.33$ & $99.67$/$\mathbf{100}$ & $91.67$ & $99.33$/$\mathbf{100}$ & $75.67$ & $98.67$/$\mathbf{100}$ & $59.67$ \\
    & $30$ & $\mathbf{100}$ & $\mathbf{100}$ & $\mathbf{100}$ & $99.67$ & $\mathbf{100}$ & $96.33$ & $\mathbf{100}$ & $89.67$ & $99.67$/$\mathbf{100}$ & $74.67$ & $98.33$/$\mathbf{100}$ & $58.33$ \\
    \midrule
    \multirow{4}{*}{$6$} & $15$ & $\mathbf{100}$ & $\mathbf{100}$ & $\mathbf{100}$ & $99.00$ & $\mathbf{100}$ & $96.67$ & $\mathbf{100}$ & $95.33$ & $\mathbf{100}$ & $89.33$ & $99.33$/$\mathbf{100}$ & $79.33$ \\
    & $20$ & $\mathbf{100}$ & $\mathbf{100}$ & $\mathbf{100}$ & $99.00$ & $\mathbf{100}$ & $97.33$ & $\mathbf{100}$ & $96.00$ & $99.67$/$\mathbf{100}$ & $84.67$ & $98.00$/$\mathbf{100}$ & $74.67$ \\
    & $25$ & $\mathbf{100}$ & $99.67$ & $\mathbf{100}$ & $99.33$ & $\mathbf{100}$ & $99.67$ & $\mathbf{100}$ & $95.67$ & $99.00$/$\mathbf{100}$ & $86.67$ & $97.00$/$\mathbf{100}$ & $68.33$ \\
    & $30$ & $\mathbf{100}$ & $99.00$ & $\mathbf{100}$ & $99.00$ & $\mathbf{100}$ & $97.00$ & $\mathbf{100}$ & $96.33$ & $99.00$/$\mathbf{100}$ & $85.33$ & $93.67$/$\mathbf{100}$ & $69.00$ \\
    \bottomrule
  \end{tabular}}}
  \caption{Solve rate (\%) on the full size$\times$depth suite. The table extends Table~\ref{tab:capdepth} to $L=200$ and $250$ and reports rates to two decimal places. The evaluation protocol and notation are as in Table~\ref{tab:capdepth}. $^{\dagger}$~At $L=200$ and $250$, the fractions of four-point inputs within both checkpoint limits are $14\%$ and $9\%$, respectively.}
  \label{tab:capdepthext}
\end{table}

\begin{table}[!htbp]
  \centering
  \footnotesize
  \setlength{\tabcolsep}{3pt}
  {\resizebox{\textwidth}{!}{%
  \begin{tabular}{l@{\hspace{6pt}}rrrr@{\hspace{12pt}}rrrr@{\hspace{12pt}}rrrr}
    \toprule
    & \multicolumn{4}{c}{$n=4$} & \multicolumn{4}{c}{$n=5$} & \multicolumn{4}{c}{$n=6$} \\
    \cmidrule(lr){2-5}\cmidrule(lr){6-9}\cmidrule(lr){10-13}
    $L$ & median & p90 & p99 & max & median & p90 & p99 & max & median & p90 & p99 & max \\
    \midrule
    $60$ & $1{,}085$ & $1{,}996$ & $6{,}990$ & $29{,}268$ & $2{,}330$ & $4{,}737$ & $9{,}610$ & $19{,}371$ & $3{,}526$ & $7{,}173$ & $12{,}964$ & $16{,}738$ \\
    $90$ & $1{,}522$ & $2{,}864$ & $11{,}467$ & $165{,}546$ & $4{,}223$ & $8{,}005$ & $15{,}095$ & $34{,}738$ & $6{,}159$ & $11{,}649$ & $19{,}118$ & $26{,}605$ \\
    $120$ & $1{,}931$ & $3{,}659$ & $15{,}819$ & $150{,}663$ & $6{,}732$ & $12{,}740$ & $25{,}300$ & $46{,}269$ & $10{,}150$ & $17{,}030$ & $28{,}134$ & $38{,}468$ \\
    $150$ & $2{,}259$ & $4{,}007$ & $15{,}032$ & $558{,}124$ & $8{,}846$ & $16{,}096$ & $25{,}190$ & $77{,}236$ & $13{,}486$ & $22{,}342$ & $33{,}574$ & $41{,}734$ \\
    $200$ & $2{,}699$ & $4{,}473$ & $11{,}018$ & $2{,}958{,}920$ & $13{,}633$ & $24{,}272$ & $40{,}404$ & $122{,}420$ & $20{,}964$ & $34{,}433$ & $53{,}985$ & $77{,}971$ \\
    $250$ & $3{,}060$ & $5{,}237$ & $15{,}622$ & $28{,}971{,}674$ & $18{,}268$ & $33{,}576$ & $63{,}393$ & $2{,}892{,}468$ & $28{,}490$ & $46{,}918$ & $71{,}438$ & $89{,}189$ \\
    \midrule
    all & $1{,}996$ & $4{,}127$ & $13{,}010$ & $28{,}971{,}674$ & $7{,}610$ & $20{,}873$ & $39{,}943$ & $2{,}892{,}468$ & $11{,}532$ & $31{,}045$ & $54{,}053$ & $89{,}189$ \\
    \bottomrule
  \end{tabular}}}
  \caption{Moves per case solved by \gbfs\ on the size$\times$depth suite, including searches with increased budgets. Statistics are marginalized over the four generation depths $K$, with $N=1{,}200$ cases for each $(n,L)$. For $(n,L)=(4,150)$, the statistics cover $1{,}199$ solved cases. These costs exclude unsuccessful attempts and separate raw-ordering searches.}
  \label{tab:budgetapps}
\end{table}

\bibliographystyle{JHEP}
\bibliography{refs}

\end{document}